\documentclass[aps,prx,twocolumn,amsfonts,floatfix,superscriptaddress]{revtex4-2}
\usepackage[utf8]{inputenc}
\usepackage{hyperref}
\def\doi{http://dx.doi.org/}
\usepackage{amsmath,braket,amssymb,xcolor,calligra,color,empheq,amsthm}
\usepackage{color}
\usepackage{epsfig}
\usepackage{tikz}
\usetikzlibrary{spy}
\usetikzlibrary{patterns}
\usetikzlibrary{calc,arrows,positioning}
\usetikzlibrary{arrows,shadows}
\usetikzlibrary{decorations.pathmorphing}	
\usetikzlibrary{decorations.markings}
\usepackage{pgfplots}
\usepackage{pgfplotstable}
\pgfplotsset{compat=1.17} 
\usepackage{subcaption}
\usepackage{float}

\def\sfix#1{\texorpdfstring{#1}{Lg}}

\newcommand{\be}{\begin{equation}}
\newcommand{\ee}{\end{equation}}
\newcommand{\bea}{\begin{eqnarray}}
\newcommand{\eea}{\end{eqnarray}}

\def\nn{\nonumber\\}

\def\fr#1{(\ref{#1})}

\def\nn{\nonumber\\}

\def\fr#1{(\ref{#1})}

\def\bla{\boldsymbol{\lambda}}
\def\bmu{\boldsymbol{\mu}}

\usepackage[most]{tcolorbox}
\tcbuselibrary{theorems}

\newtcbtheorem{Aside}{Remark}%
{breakable,colframe=blue!50!white!50!black,fonttitle=\bfseries}{}

\newtcbtheorem{Proposition}{}%
{breakable,colframe=red!50!white!50!black,fonttitle=\bfseries}{}

\begin{document}

\title{Statistics of matrix elements of local operators in integrable models}

\author{F.H.L. Essler} 
\affiliation{Rudolf Peierls Centre for Theoretical Physics, Clarendon
  Laboratory, Oxford OX1 3PU, UK}
\email{fab@thphys.ox.ac.uk}

\author{A.J.J.M. \surname{de Klerk}}
\affiliation{Institute for Theoretical Physics, University of Amsterdam,
Postbus 94485, 1090 GL Amsterdam, The Netherlands}

\begin{abstract}
We study the statistics of matrix elements of local operators in the
basis of energy eigenstates in a paradigmatic integrable many-particle
quantum theory, the Lieb-Liniger model of bosons with repulsive
delta-function interaction. Using methods of quantum
integrability we determine the scaling of matrix elements with system
size. As a consequence of the extensive number of 
conservation laws the structure of matrix elements is fundamentally
different from, and much more intricate than, the predictions of the
eigenstate thermalization hypothesis for generic models. We uncover an
interesting connection between this structure for local operators
in interacting integrable models, and the one for local operators that
are not local with respect to the elementary excitations in free
theories. We find that typical off-diagonal matrix elements
$\langle\boldsymbol{\mu}|{\cal O}|\boldsymbol{\lambda}\rangle$
in the same macro-state scale as $\exp(-c^{\cal O}L\ln(L)-LM^{\cal
  O}_{\boldsymbol{\mu},\boldsymbol{\lambda}})$ where the probability
distribution function for $M^{\cal
  O}_{\boldsymbol{\mu},\boldsymbol{\lambda}}$ are well described by 
Fr\'echet distributions and $c^{\cal O}$ depends only on macro-state
information. In contrast, typical off-diagonal matrix elements
between two different macro-states scale as $\exp(-d^{\cal O}L^2)$,
where $d^{\cal O}$ depends only on macro-state information. 
Diagonal matrix elements depend only on macro-state information
up to finite-size corrections.
\end{abstract}

\date{\today}

\maketitle
\tableofcontents
\section{Introduction}
To fully characterize the mechanism that underlies the emergence of
equilibrium statistical mechanics from the non-equilibrium evolution of
many-particle quantum systems has been a long standing challenge in
theoretical physics. A key element of our current understanding is
the Eigenstate Thermalization Hypothesis (ETH)\cite{deutsch1991quantum,srednicki1994chaos,srednicki1999approach,dalessio2016quantum}, which relates
thermalization in ``generic'' quantum  systems to the statistical
properties of matrix elements of (local) operators in energy
eigenstates. Here the term ``generic'' refers in particular to the
absence of conservation laws with local densities other than the
energy itself. The ETH is a conjecture for the matrix elements ${\cal
  O}_{nm}=\langle n|{\cal O}|m\rangle$ in the energy eigenbasis
$H|n\rangle=E_n|n\rangle$ and reads
\be
{\cal
  O}_{nm}=O(\bar{E})\delta_{n,m}+e^{-\frac{1}{2}S(\bar{E})}f_{\cal
  O}(\bar{E},\omega)R_{nm}\ .
\label{ETH}
\ee
Here $\bar{E}=(E_n+E_m)/2$, $\omega=E_n-E_m$, $S(\bar{E})$ is the
thermodynamic entropy at energy $\bar{E}$, $R_{nm}$ are random
variables with zero mean and unit variance, and
$O(\bar{E})$ and $f_{\cal O}(\bar{E},\omega)$ are smooth functions of
their arguments. The ETH implies that time averages of observables
after a quantum quench from an initial state with sub-extensive energy
fluctuations converge to a steady state, which is equivalent to the
micro-canonical ensemble. The ETH conjecture is consistent with
numerous numerical studies 
\cite{rigol2008thermalization,rigol2010quantum,steinigeweg2013eigenstate,kim2014testing,beugeling2014finitesize,beugeling2015offdiagonal,chandran2016the,mondani2017eigenstate,nation2018offdiagonal,yoshizawa2018numerical,khaymovich2019eigenstate}. A
recent focus has been to clarify the statistical properties of the
random variables $R_{nm}$ \cite{pappalardi2022eigenstate}. By
construction ETH only applies to generic models and needs to be
modified in the presence of conservation laws. In particular, it clearly
does not hold in non-interacting theories. This in turn generated
significant interest \cite{biroli2010effect,ikeda2013finitesize,alba2015eigenstate,khatami2013fluctuation,leblond2019entanglement,brenes2020lowfrequency,leblond2020eigenstate,mierzejewski2020quantitative,zhang2022statistical}
in the question of what takes the place of the
ETH in integrable models
\cite{korepin1993quantum,takahashi1999thermodynamics,essler2005one,gaudin2014bethe},
which are characterized by having an extensive number of mutually
compatible conserved quantities with good spatial locality
properties. Curiously, most studies of the statistics of matrix
elements in integrable models have not utilized the available analytic
results of the structure of these matrix elements
\cite{smirnov1992form,korepin1993quantum,korepin1982calculation,slavnov1989calculation,caux2007one,piroli2015exact},
and as a result have been restricted to very small system sizes/low
particle numbers. This has in particular precluded a study of how
matrix elements scale with system size, which is a serious shortcoming
as one is of course ultimately interested in understanding how the 
thermodynamic limit is approached. The purpose of this work is to
fully utilize the available information from integrability in order to
understand the statistic of matrix elements of local operators in
interacting integrable models. We will focus on a particular model --
the Lieb-Liniger model of bosons with delta-function interactions
\cite{lieb1963exact,korepin1993quantum} -- but we believe our results
to carry over to other integrable models. Our choice is based on the
following two requirements: 
\begin{enumerate}
\item{} We must be able to compute matrix elements for large systems
sizes/particle numbers for energy eigenstates at finite energy
densities above the ground state;
\item{} We seek an integrable model with free parameters that is
equivalent to a non-interacting theory at particular points in
parameter space.
\end{enumerate}
Among these the first point is a much more serious restriction. Almost
all integrable models feature hierarchies of multi-particle bound
states called ``strings''
\cite{bethe1931theorie,korepin1993quantum,takahashi1999thermodynamics,essler2005one,gaudin2014bethe},
and it is well-understood that the most prevalent (thermal) states at
a given energy density involve finite densities of strings. On the
one hand this makes sampling such states in a large finite volume
very challenging, but more importantly, the corresponding matrix
elements become highly singular already for very moderate system sizes
and their numerical evaluation remains an unsolved problem. This in
turn means that for integrable models like the spin-1/2 XXZ chain
matrix elements involving thermal states, i.e. the most likely states
at a given energy density, cannot be computed for large system
sizes\slash particle numbers. We avoid this issue by focusing on the
repulsive Lieb-Liniger model, where bound states are absent and matrix
elements involving thermal states can be readily investigated for
large system sizes.

\subsection{The Lieb-Liniger Model}
The Lieb-Liniger model of bosons with $\delta$-function
interaction~\cite{lieb1963exact,korepin1993quantum} is described by
the second-quantized Hamiltonian 
\be
H=\int {\rm d}x \Big(
-\Phi^\dagger(x)\partial_x^2\Phi(x)+ c \big(\Phi^\dagger(x)\big)^2\big(\Phi(x)\big)^2 \Big)\ ,
\label{hamLL}
\ee
where $\Phi(x)$ is a complex bosonic field obeying canonical commutation relations
\be
[\Phi(x),\Phi^\dagger(y)]=\delta(x-y).
\label{CCR}
\ee
The Hamiltonian has a U(1) symmetry related to particle number
conservation and its first quantized form in the $N$-particle sector
reads 
\be
\hat{H} = \sum_{j=1}^N - \frac{\partial}{\partial x_j^2} + 2c \sum_{i<j} \delta(x_i-x_j).
\label{H1}
\ee
The Lieb-Liniger model is not only a key paradigm for integrable
many-particle quantum models \cite{korepin1993quantum}, but has been
(approximately) realized in cold atom experiments, see e.g. the
reviews~\cite{bloch2008manybody,cazalilla2011one}. This motivated
an intense effort in recent years aimed at understanding dynamical
properties of the model both in 
\cite{caux2007one,kitanine2012form,fabbri2015dynamical,meinert2015probing,kozlowski2015large,doyon2017drude,doyon2018exact,granet2020a,granet2021low}
and out of equilibrium \cite{caux2013time,kormos2013interaction,denardis2014solution,kormos2014analytic,denardis2015density,piroli2016multiparticle,bouchoule2022generalized,robinson2021on,granet2021systematic}.

The outline of this work is as follows. In
Sec.~\ref{sec:EnergyEigenstates} we briefly review some important 
properties of energy eigenstates in integrable models. In particular
we introduce the notion of macro-states as families of energy
eigenstates characterized by the same densities of the conservation
laws, which lies at the heart of our analysis of matrix elements. In
section \ref{sec:generating} we discuss how to efficiently sample
energy eigenstates belonging to a given macro-state. This is crucial
as the total number of energy eigenstates grows exponentially with
particle number if we impose a momentum cutoff.
In section \ref{sec:Operators} we introduce the operators whose matrix
elements we consider in this work, and introduce the notion of
locality of an operator relative to the elementary excitations of the
model considered. In sections \ref{sec:MEfree} and
\ref{sec:Offdiagonalfree} we analyze matrix elements in free
theories by considering the example of the impenetrable Bose gas
$c=\infty$. While these are simple for operators that are local with 
respect to the elementary excitations, we reveal an intricate
structure of matrix elements of the Bose field, which is the simplest
example of a local operator that is not local with respect to the
fermionic elementary excitation. In section \ref{sec:MEint} we then
turn to the statistics of matrix elements in the interacting case
$0<c<\infty$ and show that their qualitative behaviour is the same as
the one we found for the Bose field in the impenetrable limit,
i.e. local operator in free theories that are not local with respect to the
elementary excitation. We summarize our results in
\ref{sec:Conclusions}. Various technical aspects of analytic
calculations and methods for sampling eigenstates are presented in two
appendices. 
\section{Energy eigenstates in integrable models}
\label{sec:EnergyEigenstates}
As we are concerned with properties of energy eigenstates in integrable models
we begin by recalling their construction in both free and interacting
theories. We draw particular attention to the thermodynamic limit
description in terms of macro-states, and how these are related to
energy eigenstates in very large systems. By virtue of the presence of an
extensive number of conservation laws these have a much more
complicated structure than in the generic models to which the ETH applies.
Our discussion follows Refs~\cite{essler2019chapter,essler2022short}. 

\subsection{Free theories}
\label{ss:freetheories}
Free (non-interacting) theories are the simplest integrable models. In
order to be as close as possible to the interacting theory discussed
later on, we focus on the example of the \emph{impenetrable} Bose gas
\cite{korepin1993quantum}, i.e. the limit $c\to\infty$ in
\fr{hamLL}. This is well known to be equivalent to a theory of free
fermions \cite{creamer1980quantum} by the mapping 
\be
\Phi^\dagger(x)=\Psi^\dagger(x)\ e^{i\pi\int_{-\infty}^x dz\ \Psi^\dagger(z)\Psi(z)}\ ,
\label{BoseFermi}
\ee
where $\Psi(x)$ is a complex fermion field obeying canonical
anti-commutation relations
$\{\Psi(x),\Psi^\dagger(y)\}=\delta(x-y)$. The second-quantized
Hamiltonian then becomes block-diagonal in the sectors with even/odd
fermion number and each block takes the simple form 
\be
H(\infty)=-\int dx\ \Psi^\dagger(x)\partial_x^2\Psi(x)\ .
\ee
The wave functions of energy eigenstates of the bosonic and fermionic
realisations are related by the celebrated Girardeau formula
\cite{girardeau1960relationship} 
\be
\chi_F(z_1,\dots,z_N)=\prod_{i<j}{\rm sgn}(z_j-z_i)\chi_B(z_1,\dots,z_N).
\ee
Having in mind this simple relationship we will therefore focus on the
construction of energy eigenstates in the free fermion
representation. The Hamiltonian on a ring of circumference $L$ is
diagonalized by going to Fourier space 
\be
H(\infty)=\sum_pp^2\Psi^\dagger_p\Psi_p\ ,
\ee
where $p=2\pi I_n/L$ with $I_n$ half-odd integers (integers) in the
sector with even (odd) fermion number and
\be
\Psi_p=\frac{1}{\sqrt{L}}\int_0^L dx\ e^{-ipx}\Psi(x)\ .
\ee
There is an extensive number of mutually compatible conservation laws
with local densities
\be
Q^{(n)}=\sum_p p^n\ \Psi^\dagger_p\Psi_p\ ,\ [Q^{(n)},Q^{(m)}]=0.
\ee
A complete set of simultaneous N-particle eigenstates of all the
$Q^{(n)}$ is given by the momentum-space Fock states 
\be
|\boldsymbol{p}\rangle=\prod_{j=1}^N\Psi^\dagger_{p_j}|0\rangle\ ,\quad p_1<p_2\dots<p_N\ ,
\ee
which have eigenvalues
\be
Q^{(n)}|\boldsymbol{p}\rangle=\sum_{j=1}^Np_j^n|\boldsymbol{p}\rangle.
\label{EVQ}
\ee
\subsubsection{Macro-states}
\label{ss:freemacro-states}
Local properties in the thermodynamic limit 
\be
N,L\to\infty\ ,\ D=\frac{N}{L}\ \text{fixed},
\ee
are conveniently described in terms of \textit{macro-states}. These
are families of energy eigenstates, which have the same local
properties. The latter are in turn fully encoded in the extensive
parts of the  eigenvalues \fr{EVQ} of the conservation laws.
These observations lead us to consider families of Fock states
$\{|k_1,\dots,k_N\rangle\}$ for asymptotically large $L$ and $N=DL$
that are characterised by a positive function $0\leq\rho(k)\leq 1$
termed the \textit{root density} through 
\be
L\rho(k)\Delta k = \text{ number of }k_j\text{ in }[k,k+\Delta k]\ .
\label{macmic}
\ee
It is then straightforward to see that any micro-state
$\{|k_1,\dots,k_N\rangle\}$ associated with the same root density
$\rho(\lambda)$ has the same extensive parts of the  eigenvalues
\fr{EVQ} of the conservation laws $Q^{(n)}$
\begin{align}
\frac{1}{L}\sum_{j=1}^Nk_j^n
=\int_{-\infty}^\infty dk\ \rho(k)\ k^n+o(L^0).
\end{align}

\paragraph{Counting microstates.}
For a given $\rho(k)$ there are generally exponentially many (in the
system size $L$) eigenstates satisfying~\fr{macmic}. In the interval
$[k,k+\Delta k]$, a momentum $k_j$ can take $\Delta n_{\rm vac} =
\lfloor L \Delta k/2\pi \rfloor$ possible values (here $\lfloor x
\rfloor$ denotes the integer part of 
$x$). The root density sets how many of these ``vacancies'' (possible
values) are occupied, with the occupation number given by $\Delta
n_{\rm p} = [ \rho(k) L \Delta k ]$. The $\Delta n_{\rm p}$ occupied
momenta can be distributed over the $\Delta n_{\rm vac}$ vacancies in  
$C(\Delta n_{\rm vac},\Delta n_{\rm p})$ possible ways, where $C(n,m)$
denotes a binomial coefficient. The entropy of our macro-states is given by
$S=\ln(\#\text{ of micro-states})$, where reordering of momenta in a
given interval $[k,k+\Delta k]$ contributes  
$\Delta S = \ln[ C(\Delta n_{\rm vac},\Delta n_{\rm p})]$.
Using Stirling's approximation under the assumption that $\Delta n_\text{vac}$ and $\Delta n_p$ scale with $L$ we then have in the large volume limit
\begin{widetext}
\bea
S[\rho] &=& s[\rho]L 
= -L\int_{-\infty}^{\infty}{\rm d} k 
\left[\big(\rho(k)+\rho_h(k)\big)\ln\big(\rho(k)+\rho_h(k)\big)
-\rho(k)\ln\big(\rho(k)\big)
-\rho_h(k)\ln\big(\rho_h(k)\big)
\right]+o(L).
\label{entropy}
\eea
\end{widetext}
Here we have defined a \textsl{hole density} by
\be
\rho_h(k)=\frac{1}{2\pi}-\rho_p(k)\ ,
\ee
\paragraph{Typical vs atypical states.}
Let us consider energy eigenstates at energy density $e$ and particle
density $D$. Clearly there will be infinitely many macro-states
satisfying these conditions: all we require is a positive function
$\rho(k)$ such that 
\begin{align}
    e=\int_{-\infty}^\infty dk\ \rho(k)\ k^2\ ,\quad
    D=\int_{-\infty}^\infty dk\ \rho(k)\ .
\label{eD}
\end{align}
Generically these macro-states will have finite entropy densities in
the thermodynamic limit, see Eq.~\fr{entropy}, and importantly these
macro-states will generally \textit{not be thermal}. Indeed, thermal
macro-states are obtained by maximising the free energy per site:  
\be
f[\rho]=\int_{-\infty}^\infty{\rm d}k\ (k^2-\mu)\rho(k)-Ts[\rho]\ ,
\ee
where $\mu$ is a chemical potential that determines $D$. This leads to
the root density taking the form of a Fermi distribution at
temperature $T$ 
\be
\frac{\delta f[\rho]}{\delta\rho(k)}=0\Longrightarrow \rho_{\rm  th}(k) = \frac{1}{2\pi}\frac{1}{e^{(k^2-\mu)/T}+1} \ .
\label{rhothermal}
\ee
Fixing the chemical potential and temperature by inserting
\fr{rhothermal} into \fr{eD} provides us with a root density of
thermal states. By construction thermal states are maximal entropy
states for given $e$ and $D$, i.e. they are the \textit{most likely
  states}. As we have seen above, other macro-states will exist at the
same energy density with entropies that are smaller than those of the
thermal state. If at a given energy density we select a micro-state at
random, this will be thermal with a probability that is exponentially
close (in system size) to one. We call such states ``typical'', while
noting that there are exponentially many micro-states that are
``atypical'', which differ from thermal micro-states in the values of
the higher conservation laws $Q^{(n)}$ and hence have different local
properties (as the densities of $Q^{(n)}$ are local operators and
macro-states are homogeneous).  

The situation with ``typical'' and ``atypical'' micro-states
generalizes to the case of integrable models with interactions
\cite{essler2019chapter}, and atypical states within these models can
have very interesting properties (see, for example,
Refs.~\cite{piroli2016multiparticle,veness2017quantum,denardis2018edge}.
\subsection{Interacting theories: Lieb-Liniger at \sfix{$0<c<\infty$}}
\label{sec:LL}
The Lieb-Liniger model is famously solvable by coordinate Bethe ansatz
\cite{lieb1963exact}, and we now briefly summarize the key steps
following Ref.~\cite{korepin1993quantum}. The
eigenvalue equation for the first quantized Hamiltonian \fr{H1} reads 
\be
 \hat H \chi(x_1,\ldots,x_N) = E \chi(x_1,\ldots,x_N)\ ,
\ee
where the wave functions fulfil periodic boundary conditions
\be
\chi_{\boldsymbol{\lambda}}(x_1,\dots ,
x_j+L,x_N)=\chi_{\boldsymbol{\lambda}}(x_1,\dots x_N)\ .
\ee
The (unnormalized) solutions take Bethe ansatz form
\bea
\chi_{\boldsymbol{\lambda}}(x_1,\dots
x_N)=\sum_{P\in S_N}&&{\rm
  sgn}(P)e^{i\sum_{j=1}^N\lambda_{P_j}x_j}\nn
  \times && \prod_{j>k}\big[\lambda_{P_j}-\lambda_{P_k}-ic\big]\ ,
\label{wavefn}
\eea
where the rapidities $\boldsymbol{\lambda} = \{ \lambda_1 , \ldots, \lambda_N\}$ satisfy non-trivial quantisation conditions known as \textit{Bethe equations}
\be
e^{i\lambda_j
  L}=-\prod_{k=1}^N\frac{\lambda_j-\lambda_k+ic}{\lambda_j-\lambda_k-ic}\ ,\quad
j=1,\dots,N.
\label{BAE1}
\ee
The energy and momentum eigenvalues of these states are
\be
E_{\boldsymbol{\lambda}}=\sum_{j=1}^N\lambda_j^2\ ,\quad
P_{\bmu}=\sum_{j=1}^N\lambda_j.
\ee
The states are in fact simultaneous eigenstates of an infinite number
of mutually compatible higher conservation laws $Q^{(n)}$
\cite{davies2011higher,korepin1993quantum} 
\begin{align}
Q^{(n)}\chi_{\boldsymbol{\lambda}}(x_1,\dots,x_N)&=\nu^{(n)}_{\boldsymbol{\lambda}}\chi_{\boldsymbol{\lambda}}(x_1,\dots,x_N)\ ,\nn
\nu^{(n)}_{\boldsymbol{\lambda}}&=\sum_{j=1}^N\lambda_j^n\ .
\label{charges}
\end{align}

In practice, we will use deal with a set of equations known as the \textit{logarithmic Bethe equations}, which are obtained by taking the logarithm of~\fr{BAE1}: 
\bea
\lambda_jL+\sum_{k=1}^N\theta(\lambda_j-\lambda_k)
=2\pi I_j\ ,\label{BAE2} \\ 
\theta(x)=2\ {\rm arctan}\bigg(\frac{x}{c}\bigg)\ . \nonumber
\eea
In taking the logarithm we introduce $I_j$, which are integers (half-odd integers) for $N$ odd (even). Each solution of the Bethe equations~\fr{BAE1} is in one-to-one correspondence with a set of distinct (half-odd) integers $\{I_j\}$, and hence the set of distinct integers defines a wave function $\chi_{\boldsymbol{\lambda}}(x_1,\dots,n_N)$ that is a simultaneous eigenstate of the Hamiltonian and the conservation laws. 
\subsubsection{Solutions of the Bethe equations}
An important simplification that occurs for the Lieb-Liniger model is
that all solutions to the Bethe equations are in fact
real \cite{korepin1993quantum}. This greatly simplifies the task of
solving the Bethe equations numerically. In other interacting
integrable models the solutions are typically complex, and form
regular patterns known as "strings"
\cite{takahashi1999thermodynamics,essler2005one}. As noted above,
solutions of the Bethe equations involving strings are numerically
very difficult to obtain, because some of the differences between the
corresponding rapidities lie exponentially (in system size) close to
poles of the Bethe equations.  
\subsubsection{Macro-states}
Given the above description of energy eigenstates in terms of the solutions of
the Bethe equations we now turn to the construction of
macro-states. The main complication here, as compared to the process
for the free theory (described in Sec.~\ref{ss:freemacro-states}), is
that the quantization conditions described in Eqs.~\fr{BAE1}
and~\fr{BAE2} are non-trivial and so the set of rapidities
$\boldsymbol{\lambda}$ are state dependent. We can, however, get
around this complication by instead working with the (half-odd)
integers $\{I_j\}$ - in analogy with Eq.~\fr{macmic} we can define a
density for $\nu_j = I_j/L$ through   
\be
L\varrho(\nu)\Delta \nu=\text{ number of }\frac{I_j}{L}\text{ in
}[\nu,\nu+\Delta \nu].
\label{macmic_int}
\ee
As in the free theory, a positive function $\varrho(\nu)$ specifies a
macro-state and corresponding microstates can be constructed by
choosing $\{I_j\}$ distributed according to $\varrho(\nu)$. In practice,
it is useful to have a formulation in terms of the distribution
function $\rho(\lambda)$ --  called root density -- of the rapidities
$\lambda_j$ that satisfy Eq.~\fr{BAE1}, defined via  
\be
L\rho(\lambda)\Delta \lambda=\text{ number of }\lambda_j\text{ in
}[\lambda,\lambda+\Delta \lambda].
\label{macmic_rap}
\ee
The relationship between $\rho(\lambda)$ and $\varrho(\nu)$ can be
obtained from Eq.~\fr{BAE2} by converting the sum over rapidities to
an integral over $\rho(\lambda)$ in the thermodynamic limit 
\bea
z_j=\frac{I_j}{L}&=&\frac{\lambda_j}{2\pi}+ \frac{1}{2\pi L}\sum_{k=1}^N \theta(\lambda_j-\lambda_k) \nn
  &\simeq&\frac{\lambda_j}{2\pi}+\frac{1}{2\pi }\int_{-\infty}^\infty {\rm d}\mu\ \theta(\lambda_j-\mu)\, \rho(\mu)\ .
\eea
Thus in the thermodynamic limit, we have 
\be
z(\lambda)=\frac{\lambda}{2\pi}+\frac{1}{2\pi }\int_{-\infty}^\infty
{\rm d}\mu\ \theta(\lambda-\mu)\, \rho(\mu)\ .
\label{count}
\ee
The strictly monotonically increasing function $z(\lambda)$ is known
as the counting function. It is useful to define a so-called
hole density $\rho_h(\lambda)$ associated to a macro-state by
taking the derivative of \fr{count}
\begin{align}
\rho(\lambda)+\rho_h(\lambda)&=
  \frac{1}{2\pi}+\int\frac{d\mu}{2\pi} K(\lambda-\mu)\, \rho(\mu)
\ ,\label{TLBAE}\\
K(\lambda)&=\frac{2c}{c^2+\lambda^2}.
\label{K}
\end{align}
The relationship between
$\varrho(z)$ and $\rho(\lambda)$ is obtained by equating the number of
rapidities and integers within each interval ${\rm d}\lambda$ 
\be
\varrho(\nu)=\rho(\lambda(\nu))\frac{d\lambda}{d\nu}.
\label{chirho}
\ee
Given that $\nu=z(\lambda(\nu))$ we have
\be
\frac{dz}{d\nu}=1=\frac{dz}{d\lambda}\frac{d\lambda}{d\nu}=
\big(\rho(\lambda)+\rho_h(\lambda)\big)\frac{d\lambda}{d\nu},
\ee
and hence
\be
\varrho(\nu)=\frac{\rho(\lambda)}{\rho(\lambda)+\rho_h(\lambda)}
=\frac{1}{1+\frac{\rho_h(z^{-1}(\nu))}{\rho(z^{-1}(\nu))}}.
\label{varrho_rho}
\ee

\subsubsection{Thermal macro-states}
Thermal macro-states are obtained by maximizing the entropy for fixed
energy and particle densities \cite{takahashi1999thermodynamics}. The
entropy density is given by the same expression \fr{entropy} as in the
non-interacting case, with the important proviso that
$\rho_h(\lambda)$ is now obtained from \fr{TLBAE}. Extremizing the
entropy for fixed energy and particle densities fixes the
corresponding root density in terms of the (nonlinear) integral equations
\begin{align}
\rho(\lambda)&=\frac{1}{2\pi(1+e^{\frac{\epsilon(\lambda)}{T}})}+
\int_{-\infty}^\infty
d\mu\ \frac{K(\lambda-\mu)}{2\pi(1+e^{\frac{\epsilon(\lambda)}{T}})}\ \rho(\mu)\ ,\nn 
\epsilon(\lambda)&=\lambda^2-h-\frac{T}{2\pi}\int d\mu\ K(\lambda-\mu)\
\ln\left[1-e^{-\frac{\epsilon(\mu)}{T}}\right].
\label{TBAeqs}
\end{align}
Here $T$ is the temperature and $h$ is a chemical potential that fixes
the particle density. The corresponding function $\varrho(\nu)$ is
obtained by determining $z(\lambda)$ from \fr{count}, and then using
\fr{varrho_rho}
\be
\varrho(\nu)=\frac{1}{1+e^{\epsilon(z^{-1}(\nu))/T}}\ .
\label{varrho_thermal}
\ee
\section{Generating micro-states for a given macro-state}
\label{sec:generating}
\begin{figure}[ht]
\includegraphics[width=0.4\textwidth]{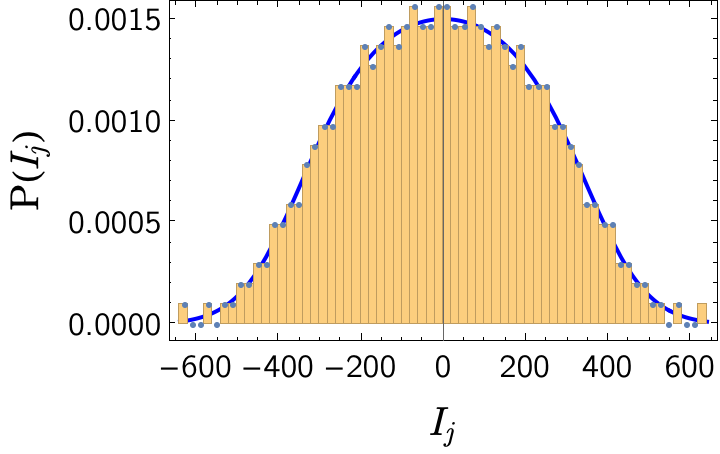}
\caption{Normalized histogram for the distribution of integers for $L=N=512$ and the "smooth" micro-state corresponding to a thermal macro-state with $c=\infty$, $\beta=0.1$ and $D=1$.}
\label{fig:intsmooth}
\end{figure}

We now turn to the problem of generating micro-states (in a large but finite
system) associated with a macro-state characterized by a root density
$\rho(x)$ in the thermodynamic limit. We start our discussion by
considering what we call "smooth" micro-states of $N$ particles in a
system of size $L$. Let us assume for definiteness that our state is
characterized by half-odd integers $I_j$. We define a "particle
counting function" by  
\be
z_p(x)=\int_{-\infty}^x dy\ \rho(y)\ ,
\ee
and then numerically solve the equations
\be
z_p(\lambda_j^{(0)})=\frac{j}{L}\ ,\ j=1,2,\dots,N.
\ee
This provides us with a set $\{\lambda_j^{(0}\}$ of rapidities. From these we generate a set of half-odd integers $I_j$ as
\be
I_j=L\ \text{Round}\Big(z(\lambda_j^{(0)})\Big) + \frac{1}{2}\text{sgn}\Big(z(\lambda_j^{(0)})\Big),
\ee
where $z(\lambda)$ is the counting function in the thermodynamic limit defined in \fr{count}.
Having determined our set of half-odd integers $\{I_1,\dots,I_N\}$ we then obtain the corresponding rapidities for a system of size $L$ by numerically solving the logarithmic form of the Bethe equations \fr{BAE2}.
The histogram of the corresponding integers or rapidities is by construction fairly smooth and closely tracks the thermodynamic root density. An example in the particularly simple case $c=\infty$ is shown in Fig.~\ref{fig:intsmooth}.
The rationale behind considering this state is that it can
be scaled up in system size, which will allow us to consider the
PDF of matrix elements between the smooth state and energy eigenstates
belonging to the same or another macro-state.

\subsection{Sampling micro-states for a given macro-state}
\label{sec:sampling}
\begin{figure}[ht]
\includegraphics[width=0.4\textwidth]{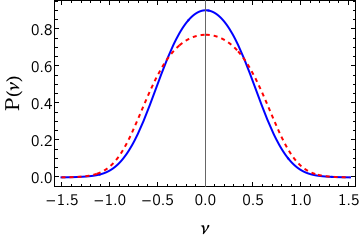}
\caption{PDF $P(\nu)$ (solid blue line) and $\varrho(\nu)$ (dashed red
  line) for $T=10$, $D=1$ and $c=\infty$.}
\label{fig:priort10}
\end{figure}

As we are interested in statistical properties of matrix elements of
local operators between energy eigenstates we require a method for
randomly sampling given classes of eigenstates. This is a necessity
because the number of micro-states corresponding to a given
macro-state grows extremely rapidly with system size, \emph{cf.} the
discussion in section \ref{sec:EnergyEigenstates}. The basic principle
is to sample the (half-odd) integers $I_j$ that specify micro-states in such a way
that they are distributed according to the distribution function
$\varrho(\nu)$ that defines the macro-state of interest. The
difficulty is knowing how close the resulting histogram for a finite
system of a few hundred particles should be to the thermodynamic
limit distribution $\varrho(\nu)$ in order for a micro-state
characterized by a set $\{I_j\}$ to ``belong'' to the macro-state
defined by $\varrho(\nu)$. A detailed discussion of this issue and its
resolution is given in Appendix \ref{app:sampling}. The upshot is that
we employ the following ``simplified random sampling'' algorithm:
\begin{enumerate}
\item{} Introduce a cutoff $I_{\rm max}$, define a set of (half-odd)
integers $\mathfrak{S}=\{-I_{\rm max},-I_{\rm max}+1,\dots,I_{\rm max}\}$ and
an empty set $S$.
\item{} Impose that the $I_j$ are distributed according to a PDF
  $P(\nu)$;
\item{} Determine the inverse $Z^{-1}(\nu)$ of the cumulative distribution function
\be
Z(\nu)=\int_0^\nu d\nu'\ P(\nu')\ ,\quad -\frac{1}{2}\leq Z(\nu)\leq\frac{1}{2}.
\ee
\item{} Generate a random number $|r|\leq \frac{1}{2}$ and use it to
generate a random integer
\be
I={\rm Round}\big(LZ^{-1}(r)\big).
\ee
\item{} We then update the sets $\mathfrak{S}$ and $S$ according to
  the rule
\begin{align}
&\text{If }I\in\mathfrak{S} \land I\notin{S}\nn
&\text{then }S\rightarrow S\cup\{I\}\ ,\
\mathfrak{S}\rightarrow\mathfrak{S}-\{I\}.
\end{align}
\item{} Repeat steps 4 and 5 until we arrive at a set
  $\{I_1,\dots,I_N\}$ of distinct integers.
\end{enumerate}
The PDF $P(\nu)$ is chosen such that this random sampling process
\emph{on average} reproduces the thermodynamic distribution
$\varrho(\nu)$. We achieve this through an iterative numerical
procedure. We show the resulting PDF for $T=10$, $D=1$ and $c=\infty$
in Fig.~\ref{fig:priort10}. 
\section{Local operators}
\label{sec:Operators}
Some representative spatially local operators in the Lieb-Liniger
model are 
\begin{itemize}
\item{Density operator}
\be
\rho(x) = \Phi^\dagger(x) \Phi(x)\ ,
\ee
\item{Interaction}
\be
g_2(x)=\big(\Phi^\dagger(x)\big)^2\big(\Phi(x)\big)^2\ ,
\label{g2x}
\ee
\item{Bose field}
\be
\Phi(x)\ .
\ee
\end{itemize}
As discussed earlier, in the limit $c\to\infty$ the model becomes equivalent to free fermions and in this limit additional local operators become of interest:
\begin{itemize}
\item{Fermi field at $c=\infty$}
\begin{align}
\Psi^\dagger(x)=\Phi^\dagger(x)\ e^{i\pi\int_{-\infty}^x dz\ \Phi^\dagger(z)\Phi(z)}\ ,
\label{BoseFermirev}
\end{align}
\item{Operator products involving the Fermi field at $c=\infty$, e.g.}
\begin{align}
J_n(x)&=(-i)^n\Psi^\dagger(x)\partial_x^n\Psi(x)\ ,\nn
F^\dagger_n(x)&=\Psi^\dagger(x)\big(\partial^n_x\Psi^\dagger(x)\big)\Psi(x)\ .
\label{BoseFermi2}
\end{align}
\end{itemize}
All of the above operators are by construction spatially local (for
finite values of $n$). However,
in integrable models the existence of stable particle and hole
excitations (at all energy densities) leads to an additional notion
of locality. The stable excitations themselves have good spatial
locality properties in the sense that they represent a local
disturbance of the macro-state under consideration. It is then natural
to ask whether a given local operator, say $\hat{\rho}(x)$, is local
with respect to the operator that creates a local stable
excitation. In relativistic integrable QFTs (at zero density) related
notions of locality are known to have far-reaching consequences for
matrix elements of local operators between energy eigenstates
\cite{smirnov1992form,lukyanov1995free}.  

In the impenetrable limit the situation becomes particularly
simple. Here the elementary excitations are fermions and are created
by $\Psi^\dagger(x)$. Operators like $\rho(x)$, $g_2(x)$,
$J_n(x)$ and of course also $\Psi^\dagger(x)$ itself are local
relative to $\Psi^\dagger(x)$ and in particular (anti)commute at a
distance. On the other hand, the Bose field itself is not local
relative to $\Psi^\dagger(x)$ as it involves a Jordan-Wigner like
string operator. As is discussed below, this leads to a dramatic
difference in the structure of matrix elements in energy eigenstates.

\section{Diagonal matrix elements of local operators in free theories}
\label{sec:MEfree}
\begin{figure}[ht]
\centering
\includegraphics[width=0.45\textwidth]{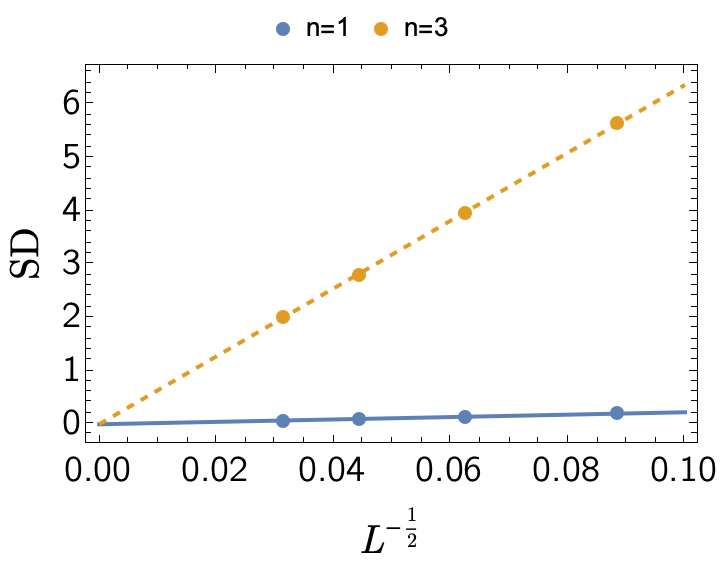}
\caption{Standard deviations of the PDFs 
$P(\mathfrak{J}_n(\boldsymbol{k}))$ for $n=1,3$
and thermal micro-states $|\boldsymbol{k}\rangle$ at $T=10$, $D=1$
sampled in a micro-canonical window $|E-e_\infty L|<10$.}
\label{fig:diagMESD}
\end{figure}

We first consider matrix elements of local operators in
non-interacting theories. The naive expectation might be that these
are trivial, but as we will see this is not the case for
operators that are not local with respect to the elementary
excitations. 
At $c=\infty$ energy eigenstates can be expressed as fermionic Fock states. Given a macro-state described by a root density $\rho(k)$ we can construct a corresponding micro-state
$|\boldsymbol{k}\rangle=|k_1,\dots,k_N\rangle$ following the procedure
outlined in section~\ref{sec:generating}. Expectation values of local
operators can then be straightforwardly calculated. Let us start with
the single-fermion Green's function at a fixed separation $x-y={\cal
  O}(L^0)$    
\begin{align}
\langle \boldsymbol{k}|\Psi^\dagger(x)\Psi(y)|\boldsymbol{k}\rangle&=
\frac{1}{L}\sum_{p,q}\langle
\boldsymbol{k}|\Psi^\dagger_p\Psi_q|\boldsymbol{k}\rangle\ e^{iyq-ipx}\nn 
&=\int_{-\infty}^\infty dk\ e^{ik(y-x)}\rho(k)+ o(L^0).
\label{expect}
\end{align}
Importantly, up to finite-size corrections this only depends on the
root density $\rho(k)$ characterizing the macro-state of interest. It
is straightforward to extend this calculation to more complicated
expectation values of the form 
\be
\langle
\boldsymbol{k}|\Psi^\dagger(x_1)\dots\Psi^\dagger(x_n)\Psi(y_n)\dots\Psi(y_1)
|\boldsymbol{k}\rangle\ , 
\ee
where we take $n$ to be fixed and all $x_j$ and all $y_\ell$ to lie in
an interval of fixed size ${\cal O}(L^0)$. Applying Wick's theorem and
using Eq.~\fr{expect} we conclude that expectation values of any
multi-point correlation function involving a fixed, finite number of
fermion operators on a finite interval can be expressed solely in
terms of the macro-state, up to finite size corrections. It then
follows in turn that expectations values of any finite number of
fermion operators calculated \textit{between different microstates
  corresponding to the same macro-state} differ only by finite-size
corrections that go to zero in the thermodynamic limit.  
Expectation values of local operators involving the Bose field at $c=\infty$ such as
\be
\langle \boldsymbol{k}|\Phi^\dagger(x)\partial_x\Phi(x)|\boldsymbol{k}\rangle
\ee
can be obtained from these results by using the Bose-Fermi mapping as the latter involves only a finite number of Fermi fields. This is in contrast to expectation values like
\be
\langle \boldsymbol{k}|\Phi^\dagger(x)\Phi(y)|\boldsymbol{k}\rangle=
\langle \boldsymbol{k}|\Psi^\dagger(x)
e^{i\pi\int_{y}^{x}dz \Psi^\dagger(z)\Psi(z)}\Psi(y)|\boldsymbol{k}\rangle\ ,
\ee
which can no longer be evaluated by using Wick's theorem for a finite
number of Fermi fields. This is intimately related to the fact that
the Bose and Fermi fields are not mutually local.  In order to assess
how quickly the diagonal matrix elements approach their thermodynamic
value with increasing $L$ we have considered the expectation values of
the operators $J_n(0)$ \fr{BoseFermi2} for $n=1,3$ in thermal
micro-states $|\boldsymbol{k}\rangle$ at temperature $T=10$ and
density $D=1$
\be
\mathfrak{J}_n(\boldsymbol{k})=
\langle\boldsymbol{k}| J_n(0)|\boldsymbol{k}\rangle
\ee
We determine the PDF of $\mathfrak{J}_n(\boldsymbol{k})$
when the thermal micro-states $|\boldsymbol{k}\rangle$ are sampled in a
micro-canonical window $|E-e_\infty L|<10$, where $e_\infty$ is the thermal
energy density in the thermodynamic limit and $E$ the energy
eigenvalue of the micro-state. The PDFs are well described by normal
distributions and their standard deviations as functions of system
size are shown in Fig.~\ref{fig:diagMESD}.
As the data is well described by simple linear fits we conclude that
the standard deviations scale to zero as $L^{-1/2}$. This is in
agreement with results on free theories in the literature
\cite{leblond2019entanglement}.
\section{Off-diagonal matrix elements in free theories}
\label{sec:Offdiagonalfree}
Our example for a free theory is again the $c=\infty$ limit of the
Lieb-Liniger model. As we will see, the structure of off-diagonal
matrix elements depends strongly on the locality properties of
local operators ${\cal O}$ relative to the Fermi field $\Psi$. As
before we consider $N$ particles on a ring of length $L$ and are
interested in the thermodynamic limit $N,L\to\infty$ at fixed $D=N/L$. 

\subsection{Local operators that are local relative to the Fermi field}
The density operator $\rho(x)$ is spatially local as well as local
with respect to the elementary fermion excitations of the Lieb-Liniger
model at $c=\infty$. Let $|\boldsymbol{\lambda}\rangle$,
$|\boldsymbol{\mu}\rangle$ be energy eigenstates with corresponding
sets of (half-odd) integers $\{I_j\}$ and $\{J_k\}$. The matrix elements
of the density operator vanish unless $\{I_j\}$ and $\{J_k\}$ differ
by precisely one particle-hole excitation 
\be
\forall j\neq a\quad I_j=J_j\ ,\quad J_a=I_a+n\notin\{I_j\}.
\ee
For such one-particle-hole excitations we have the simple result
\be
\frac{|\langle\boldsymbol{\mu}|\rho(0)|\boldsymbol{\lambda}\rangle|^2}{\langle\boldsymbol{\lambda}|\boldsymbol{\lambda}\rangle\langle\boldsymbol{\mu}|\boldsymbol{\mu}\rangle}\Bigg|_{\rm 1ph}=\frac{1}{L^2}\ .
\ee
If we introduce a cut-off $\Lambda$ in momentum the total number
$N_{\rm ph}$ of
states $|\boldsymbol{\mu}\rangle$ that lead to non-vanishing off-diagonal matrix elements
scales polynomially with system size 
\be
N_{\rm ph}=N \Big(\frac{L\Lambda}{2\pi}-N\Big).
\ee
The structure of matrix elements of other local operators that are
mutually local with the Fermi field is analogous: only a very small
fraction of all off-diagonal matrix elements are non-zero. 
\subsection{Local operators that are not local relative to the Fermi field}
As an example of a local operator that is not local relative to the
Fermi field we consider the Bose field operator, which fulfils
\be
\Phi(x)\Psi(y)=\text{sgn}(x-y)\Psi(y)\Phi(x)\ .
\ee
A convenient representation for the matrix elements of the Bose field
operator at positive values of $c$ was derived in
\cite{caux2007one}. In the impenetrable limit the matrix element between a
state $|\boldsymbol{\lambda}\rangle$ with $N$ particles and a state
$|\boldsymbol{\mu}\rangle$ is non-vanishing only if the latter
has $N-1$ particles and then reads 
\begin{align}
\frac{|\langle\boldsymbol{\mu}|\Phi(0)|\boldsymbol{\lambda}\rangle|^2}{\langle\boldsymbol{\lambda}|\boldsymbol{\lambda}\rangle\langle\boldsymbol{\mu}|\boldsymbol{\mu}\rangle}=&\frac{2^{2N-2}}{L^{2N-1}}\prod_{j=1}^N
\prod_{k=1}^{N-1}\frac{1}{(\lambda_j-\mu_k)^2}\nn
&\times\prod_{j>k}^N(\lambda_j-\lambda_k)^2\prod_{j>k}^{N-1}(\mu_j-\mu_k)^2.
\label{FFfieldcinfty}
\end{align}
This result already shows that all matrix elements compatible with
the simple particle number selection rule that $\Phi(0)$ changes
particle number by one, are non-vanishing. This is in marked contrast
to what we have for local operators that are local relative to the
Fermi field.
\subsubsection{Matrix elements involving different macro-states}
If $|\boldsymbol{\lambda}\rangle$ and $|\boldsymbol{\mu}\rangle$ belong to different macro-states, say with root densities $\rho_0(\lambda)$ and $\rho_1(\mu)$ respectively, it is straightforward to determine the leading contribution (in $L$) for large system sizes by noting that
\begin{align}
&\frac{1}{L^2}\ln\Big[\frac{|\langle\boldsymbol{\mu}|\Phi(0)|\boldsymbol{\lambda}\rangle|^2}{\langle\boldsymbol{\lambda}|\boldsymbol{\lambda}\rangle\langle\boldsymbol{\mu}|\boldsymbol{\mu}\rangle}\Big]=
-\frac{1}{L^2}\sum_{j,k}\ln\big(\lambda_j-\mu_k)^2\nn
&+\frac{1}{L^2}\sum_{j>k}^N\ln(\lambda_j-\lambda_k)^2+
\frac{1}{L^2}\sum_{j>k}^{N-1}\ln(\mu_j-\mu_k)^2+o(L^0).
\label{differentmacro}
\end{align}
Turning sums into integrals this becomes
\begin{widetext}
\begin{align}
\frac{1}{L^2}\ln\Big[\frac{|\langle\boldsymbol{\mu}|\Phi(0)|\boldsymbol{\lambda}\rangle|^2}{\langle\boldsymbol{\lambda}|\boldsymbol{\lambda}\rangle\langle\boldsymbol{\mu}|\boldsymbol{\mu}\rangle}\Big]=
\frac{1}{2}\int_{-\infty}^\infty d\lambda d\mu\ \big[\rho_0(\lambda)-\rho_1(\lambda)\big]
\big[\rho_0(\mu)-\rho_1(\mu)\big] \ln\big(\lambda-\mu)^2
+o(L^0).
\end{align}
\end{widetext}
This tells us that matrix elements involving two different macro-states are extremely small
\be
\frac{|\langle\boldsymbol{\mu}|\Phi(0)|\boldsymbol{\lambda}\rangle|^2}{\langle\boldsymbol{\lambda}|\boldsymbol{\lambda}\rangle\langle\boldsymbol{\mu}|\boldsymbol{\mu}\rangle}\propto e^{-c_{\rho_0,\rho_1}L^2}\ .
\label{offdiagfree}
\ee
This behaviour is in stark contrast to the behaviour of off-diagonal
matrix elements in different macro-states non-integrable models as
predicted by the ETH. We note that the subleading terms (in system
size) in \fr{differentmacro} depend on the details of the micro-states
$|\boldsymbol{\mu}\rangle$ and $|\boldsymbol{\lambda}\rangle$ and not
only on the macro-state information encoded in $\rho_{0,1}(\lambda)$.

\subsubsection{Typical matrix elements in the same thermal macro-state}
When $|\boldsymbol{\lambda}\rangle$ and $|\boldsymbol{\mu}\rangle$
belong to the same macro-state the leading (in $L$) term
\fr{offdiagfree} vanishes as can be seen by taking
$\rho_1(\lambda)=\rho_0(\lambda)$. 

In order to understand the structure of the subleading terms we first
fix $|\boldsymbol{\lambda}\rangle$ to correspond to a thermal state at
temperature $T=10$ and density $D=1$ and then numerically determine
the probability distribution of 
\be
\mathfrak{M}_{\boldsymbol{\lambda},\boldsymbol{\mu}}=-\frac{1}{L}\ln\Big[\frac{|\langle\boldsymbol{\mu}|\Phi(0)|\boldsymbol{\lambda}\rangle|^2}{\langle\boldsymbol{\lambda}|\boldsymbol{\lambda}\rangle\langle\boldsymbol{\mu}|\boldsymbol{\mu}\rangle}\Big]\ ,
\label{MEfrak}
\ee
where $|\boldsymbol{\mu}\rangle$ are micro-states corresponding to the
same thermal macro-state and are taken to have energy eigenvalues such
that $|E(\boldsymbol{\mu})-E(\boldsymbol{\lambda})|<25$. When
$|\boldsymbol{\lambda}\rangle$ has $L$ rapidities, the states
$|\mu\rangle$ must have $L-1$ particles in order for
$\mathfrak{M}_{\boldsymbol{\lambda},\boldsymbol{\mu}}$ to be
non-vanishing. 

\begin{figure}[ht]
\includegraphics[width=0.4\textwidth]{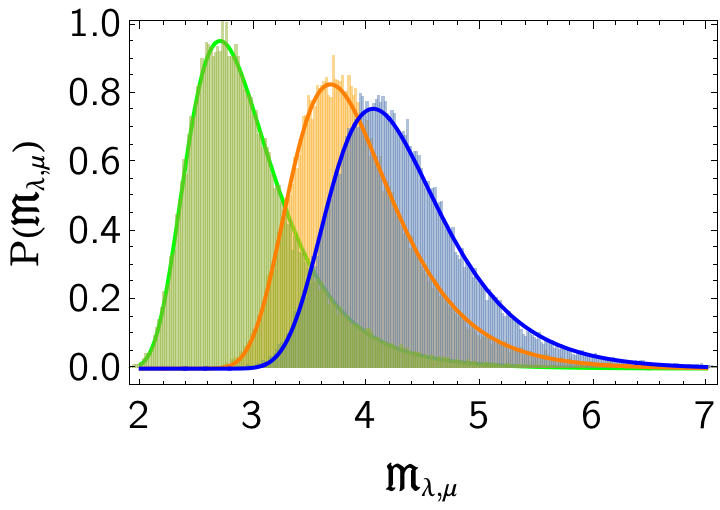}
\caption{Normalized histograms of $|M_{\boldsymbol{\lambda},\boldsymbol{\mu}}|$ for three different micro-states $|\boldsymbol{\lambda}\rangle$ (see text) and 50000 states $|\boldsymbol{\mu}\rangle$ with $L=N=512$ (respectively in yellow, blue, green, red), where $|\boldsymbol{\lambda}\rangle$ and $|\boldsymbol{\mu}\rangle$ are micro-states corresponding to the thermal macro-state at temperature $T=10$.The solid lines are fits to Fr\'echet distribution functions.}
\label{fig:state_dep}
\end{figure}

\begin{figure}[ht]
\includegraphics[width=0.4\textwidth]{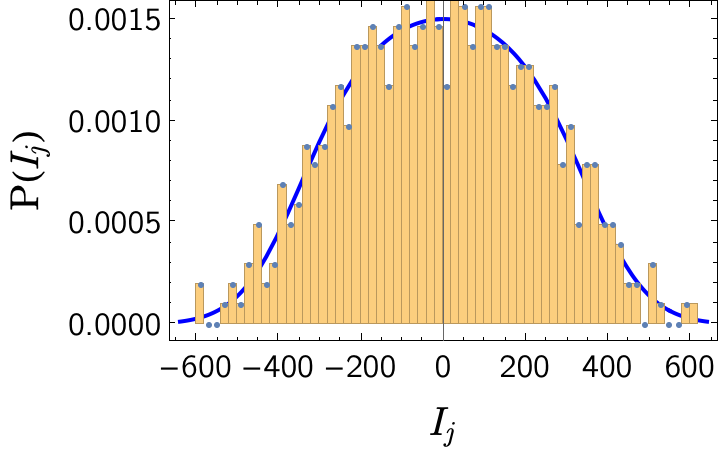}
\caption{Normalized histogram for the distribution of integers for
$L=N=512$ and a micro-state corresponding to a thermal
macro-state with $c=\infty$, $\beta=0.1$ and $D=1$} 
\label{fig:nonsmoothints}
\end{figure}

In Fig.~\ref{fig:state_dep}, where we plot the probability
distributions, obtained by sampling $\langle\boldsymbol{\mu}|$, for
three different choices of the micro-state
$|\boldsymbol{\lambda}\rangle$. Here all states belong to the same
thermal macro-state with temperature $T=10$ and  $L=N=512$.
We see that the probability distributions are very sensitive to the details of the
micro-state $|\boldsymbol{\lambda}\rangle$, and not only on
macro-state information encoded in $\rho_0(\lambda)$.
The smallest mean value of
${\mathfrak{M}}_{\boldsymbol{\lambda},\boldsymbol{\mu}}$
(corresponding to the
largest average absolute value of the matrix elements) is
obtained when $|\boldsymbol{\lambda}\rangle$ corresponds to the smooth
micro-state, \emph{cf.} the green histogram in Fig.~\ref{fig:intsmooth}. 
The yellow histogram in Fig.~\ref{fig:state_dep}, corresponding to the
second-smallest mean of the distribution, is obtained by choosing a
micro-state with the distribution of half-odd integers  
shown in Fig.~\ref{fig:nonsmoothints}. 
We see that the distribution of integers in
Fig.~\ref{fig:nonsmoothints} does not reproduce the thermodynamic root
density as well as the smooth state does. This notion can be
quantified by computing the mean-squared distance between the
histogram with bins $[\nu_1,\dots,\nu_{n_{\rm bin}+1}]$ 
\begin{align}
\Delta=\sum_{j=1}^{n_{\rm bin}} \Big[n_j-
\int_{\nu_j}^{\nu_{j+1}}d\nu \frac{2\pi}{L}\rho\Big(\frac{2\pi}{L}\nu\Big)\Big]^2.
\end{align}
Here $n_j$ is the occupation of bin $j$ and $\rho(x)$ the
thermodynamic root density describing the macro-state under
consideration. The third micro-state considered in
Fig.~\ref{fig:state_dep} (blue histogram) has the largest distance in this sense to the
thermodynamic root density. This suggests that the larger the
deviations of the root distribution of $|\boldsymbol{\lambda}\rangle$ from the
thermodynamic root density are, the smaller the typical matrix elements
$\mathfrak{M}_{\boldsymbol{\lambda},\boldsymbol{\mu}}$ (sampled over
$\langle\boldsymbol{\mu}|$) become. 

The solid lines in Fig.~\ref{fig:state_dep} are fits to Fr\'echet distribution functions 
\be
P_{\alpha,\beta,\nu}(x)=\begin{cases}
(x-\nu)^{-\alpha-1}\exp\Big[-\big(\frac{x-\nu}{\beta}\big)^{-\alpha}\Big]
&\text{if } x>\nu\\
0 &\text{else}.
\end{cases}
\ee
We find that fits to $P_{\alpha,\beta,\nu}(x)$ provide excellent
descriptions of our numerical PDFs in all cases we have considered.
The parameters $\alpha,\beta,\nu$ depend not only on macro-state
information, but on details of the micro-state
$|\boldsymbol{\lambda}\rangle$, i.e. 
\be
\alpha=\alpha_{\boldsymbol{\lambda}}\ ,\quad
\beta=\beta_{\boldsymbol{\lambda}}\ ,\quad \nu=\nu_{\boldsymbol{\lambda}}.
\label{alphabeta}
\ee

\begin{figure}[ht]
\includegraphics[width=0.4\textwidth]{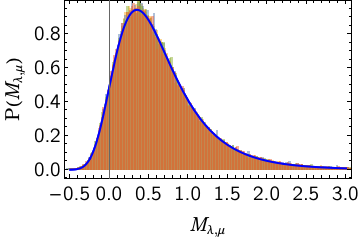}
\caption{Normalized histograms of
  $M_{\boldsymbol{\lambda},\boldsymbol{\mu}}$ for the "smooth"
  micro-state $|\boldsymbol{\lambda}\rangle$ (see text) and 50000
  states $|\boldsymbol{\mu}\rangle$ with $L=N=128,256,512,1024$
  (respectively in yellow, blue, green, red), where
  $|\boldsymbol{\lambda}\rangle$ and $|\boldsymbol{\mu}\rangle$ are
  micro-states corresponding to the thermal macro-state at temperature
  $T=10$. The solid line is a Fr\'echet distribution function with
  fitted parameters $\alpha=12.8894$, $\beta=5.04354$ and
  $\nu=-4.66742$.} 
\label{fig:collapse}
\end{figure}

The next question we want to address is how the PDFs of
$M_{\boldsymbol{\lambda},\boldsymbol{\mu}}$ scale with system size. To
address this issue we work with the smooth state, because it can be
readily scaled up with system size. We observe that we can achieve
excellent data collapse if we shift the matrix elements by a
$L$-dependent constant 
\be
{M}_{\boldsymbol{\lambda},\boldsymbol{\mu}}=
\mathfrak{M}_{\boldsymbol{\lambda},\boldsymbol{\mu}}-c_0\ln(L)\ .
\label{defM}
\ee
In Fig.~\ref{fig:collapse} we show the histograms of
${M}_{\boldsymbol{\lambda},\boldsymbol{\mu}}$ when sampled over the
states $\langle\boldsymbol{\mu}|$ for a thermal macro-state with
temperature $T=10$ and density $D=N/L=1$ for four different values of
$L$ and $c_0=0.375801$. We observe that the data for different system
sizes collapses very nicely.
Other micro-states are more difficult to scale up in system size, but
supposedly an analogous data collapse of shifted distributions occurs.  

\begin{figure}[ht]
\includegraphics[width=0.4\textwidth]{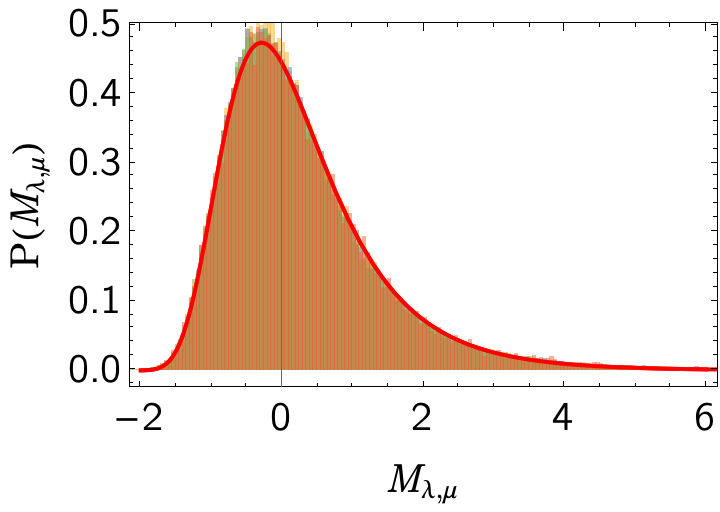}
\caption{Normalized histograms of $M_{\boldsymbol{\lambda},\boldsymbol{\mu}}$ where both  $|\boldsymbol{\lambda}\rangle$ and $|\boldsymbol{\mu}\rangle$ are sampled
from a thermal macro-state at $T=10$, $D=1$ in a fixed energy window $|E-Le_{\infty}|<7.5$
with $L=N=64,128,256,512$ (respectively in yellow, blue, green and orange).The solid line is a Fr\'echet distribution function with fitted parameters $\alpha=13.0393$, $\beta=10.1444$ and $\nu=-10.3779$.}
\label{fig:collapse2}
\end{figure}

In order to remove the explicit dependence of $P(M_{\boldsymbol{\lambda},\boldsymbol{\mu}})$
on the ket micro-state $|\boldsymbol{\lambda}\rangle$, we may sample
the latter in the same  energy window as the bra states
$\langle\boldsymbol{\mu}|$. Denoting the energy density in the 
thermodynamic limit by $e_\infty$ we take this window to be
$|E-Le_\infty|<7.5$. The resulting probability distributions of
appropriately shifted matrix elements \fr{defM} is shown in
Fig.~\ref{fig:collapse2} for a range of system sizes. 
Fixing the constant in \fr{defM} to be $c_0=0.755474$ leads to an excellent data collapse, and the resulting probability distribution is again well described by a Fr\'echet distribution.

\subsubsection{Typical matrix elements in the same non-thermal macro-state}

\begin{figure}[ht]
\includegraphics[width=0.4\textwidth]{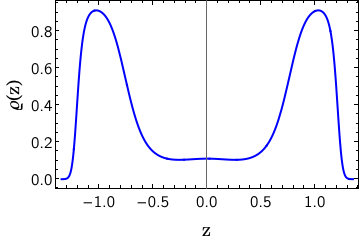}
\caption{Distribution function of $I_j/L$ for a non-thermal macro-state.}
\label{fig:probdist_atypical}
\end{figure}

\begin{figure}[ht]
\includegraphics[width=0.4\textwidth]{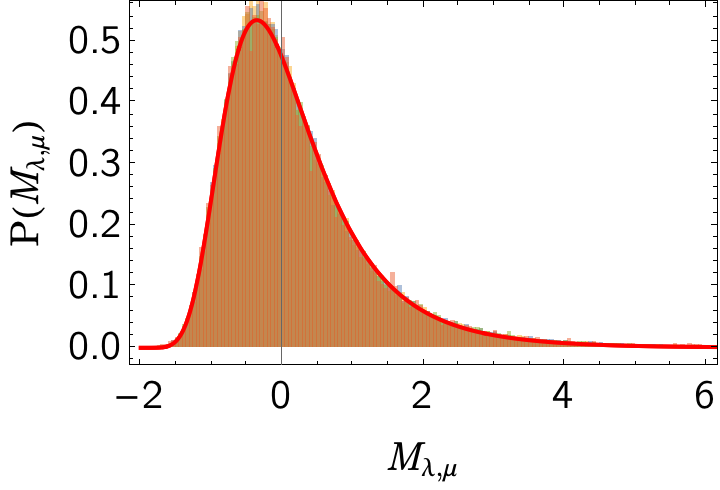}
\caption{Normalized histograms of
  $M_{\boldsymbol{\lambda},\boldsymbol{\mu}}$ where both
  $|\boldsymbol{\lambda}\rangle$ and $|\boldsymbol{\mu}\rangle$ are
  sampled from the atypical macro-state described by the root density
  \fr{rhoatypical} in fixed windows of the energy and relevant higher
  conservation laws (see text) with $L=N=64,128,192,256$ (respectively
  in yellow, blue, green, orange).The solid line is a Fr\'echet
  distribution function with fitted parameters $\alpha=8.99436$,
  $\beta=6.22114$ and $\nu=-6.50698$.} 
\label{fig:collapse3}
\end{figure}

We have also considered typical matrix elements in atypical
macro-states. As a particular example we present results for the
distribution function of integers shown in
Fig.~\ref{fig:probdist_atypical}. 
This corresponds to a generalized Gibbs ensemble with momentum distribution function
\be
\rho(\lambda)=\frac{1}{2\pi\big[1+e^{(\lambda^2-
\mu)/T+\mu_3\lambda^4+\mu_4\lambda^6}\big]}\ ,
\label{rhoatypical}
\ee
where $T=20$, $\mu=32.0846$, $\mu_3=-0.01$, $\mu_4=0.00015$.
The densities of energy and the fourth and sixth conservation law of the Lieb-Liniger model for this macro-state in the thermodynamic limit are respectively
\be
e_\infty=32.0846\ ,\ \
q^{(4)}=1270.96\ ,\ \
q^{(6)}=55027.1\ .
\ee
In order to sample the macro-state we have chosen windows for the
eigenvalues $|E-Le_\infty|<20$, $|\nu^{(4)}-Lq^{(4)}|<800$ and
$|\nu^{(6)}-Lq^{(6)}|<34300$. We note that if we do not restrict the
eigenvalues the probability distribution shifts by a small amount. The
PDF of $M_{\boldsymbol{\lambda},\boldsymbol{\mu}}$, where both
$|\boldsymbol{\lambda}\rangle$ and $|\boldsymbol{\mu}\rangle$ are
sampled from the atypical macro-state constructed in this way and the
constant shift is taken to be $c_0=0.680685$, is shown for a range of system
sizes $64\leq N\leq 256$ in Fig.~\ref{fig:collapse3}. We observe an
excellent data collapse to a PDF that is well described by a Fr\'echet
distribution function with fitted parameters $\alpha=8.99436$,
$\beta=6.22114$ and $\nu=-6.50698$.

Our results in this subsection can be summarized as follows. 
\begin{itemize}
\item{}
If we fix the ket state $|\boldsymbol{\lambda}\rangle$, then typical
off-diagonal matrix elements in the same macro-state scale with system
size as 
\be
\frac{|\langle\boldsymbol{\mu}|\Phi(0)|\boldsymbol{\lambda}\rangle|^2}{\langle\boldsymbol{\lambda}|\boldsymbol{\lambda}\rangle\langle\boldsymbol{\mu}|\boldsymbol{\mu}\rangle}\propto e^{-c_0L\ln(L)-c_1L}\ .
\label{offdiagfree2}
\ee
The corresponding probability distribution depends on details of
$|\boldsymbol{\lambda}\rangle$, i.e. the multi-variate distribution
function on (half-odd) integers and not only on macro-state
information. In particular the two constants $c_{0,1}$ depend on the
choice of $|\boldsymbol{\lambda}\rangle$. 
\item{}
The probability distribution is well fitted by a Fr\'echet
distribution, where the parameters depend on the choice of
$|\boldsymbol{\lambda}\rangle$. 
\item{} 
If we sample both the bra and ket states over the same energy window
typical matrix elements again scale with system-size as
\fr{offdiagfree2}, and the resulting probability distribution is again
well-fitted by a Fr\'echet distribution. In this case we expect
$c_{0,1}$ to depend only on macro-state information. 
\end{itemize}
An immediate consequence of the $e^{-c_0L\ln( L)}$ factor in
\fr{offdiagfree2} is that typical matrix elements will not contribute
to correlation functions of local operators in the thermodynamic
limit. To see this let us consider a two-point function in an energy
eigenstate (generalized micro-canonical ensemble, \emph{cf.}
Ref.~\onlinecite{essler2016quench}) corresponding to a macro-state
characterized by the density $\varrho_0(z)$ 
\be
F(x,t)=\langle\boldsymbol{\lambda}|\Phi^\dagger(x,t)\Phi(0,0)|\boldsymbol{\lambda}\rangle\ .
\ee
Employing a Lehmann representation and using the fact that matrix
elements involving eigenstates corresponding to different macro-states
$\varrho_1$ scale as $e^{-c_{\varrho_0,\varrho_1} L^2}$, we have
\begin{align}
F(x,t)=&\sum_{\boldsymbol{\mu}}|\langle\boldsymbol{\lambda}|\Phi^\dagger(0,0)|\boldsymbol{\mu}\rangle|^2
e^{-it
  (E_{\boldsymbol{\mu}}-E_{\boldsymbol{\lambda}})+ix(P_{\boldsymbol{\mu}}-P_{\boldsymbol{\lambda}})}\nn
&+o(L^0) \ ,
\label{fxt}
\end{align}
where the sum is over all solutions to the Bethe ansatz equations that
correspond to the macro-state characterized by $\varrho_0(z)$. Typical
matrix elements cannot contribute to this sum because they scale with
system size like  \fr{offdiagfree2}, while their number scales as
$e^{Ls_{\varrho_0}}$, where $s_{\varrho_0}$ is the thermodynamic entropy density of the
macro-state under consideration. The spectral sum \fr{fxt} must
therefore be determined by "anomalously large" matrix elements in the
"nose" of the probability distribution function. We turn to the
question of how to characterize them next. 

\subsubsection{Rare large matrix elements in the same macro-state}
\begin{figure}[ht]
\includegraphics[width=0.4\textwidth]{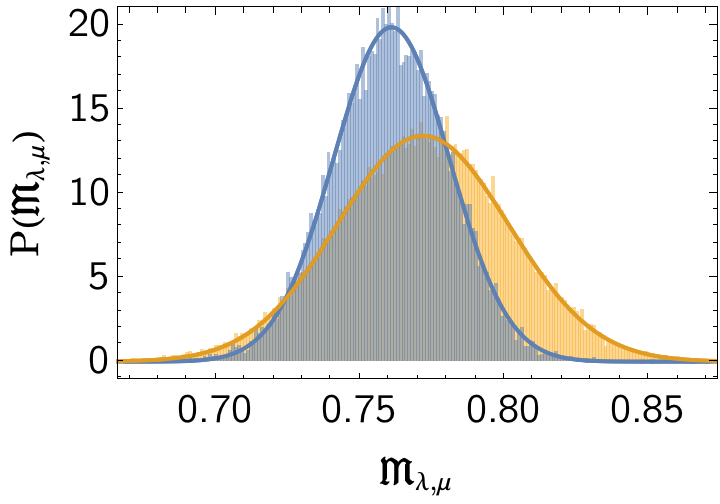}
\caption{Normalized histograms of the matrix-elements of the field
  operator between a smooth thermal state with $D=1$ and $\beta=0.1$
  and the atypical states in the set $\mathfrak{S}_0$ described in the
  text for $L=128$ (yellow)   and $L=256$ (blue). The solid lines are
  fits to normal   distributions.} 
\label{fig:atypical_free}
\end{figure}

Which states $\langle\boldsymbol{\mu}|$ give anomalously large matrix
elements
$|\langle\boldsymbol{\mu}|\Phi(0,0)|\boldsymbol{\lambda}\rangle|$
for a a given ket state $|\boldsymbol{\lambda}\rangle|$?
A natural
guess is that each of the rapidities $\mu_j$ should be very close to
one the rapidities $\lambda_k$ of the state
$|\boldsymbol{\lambda}\rangle$, so that the factor
$(\lambda_k-\mu_j)^{-2}$ in the expression of the matrix-element
\fr{FFfieldcinfty} becomes very large. This intuition is indeed
correct, as was shown for the case of the transverse-field Ising model
in Ref.~\cite{granet2020finite} (see also Refs
\cite{calabrese2012quantum,bertini2014quantum,schuricht2012dynamics}). In
Fig.~\ref{fig:atypical_free} we present histograms of matrix elements
\fr{MEfrak} for a smooth thermal state $|\boldsymbol{\lambda}\rangle$
at density $D=1$ and inverse temperature $\beta=0.1$ and a class of
states $\langle\boldsymbol{\mu}|$ selected as follows: 
\begin{itemize}
\item{} We randomly remove one of the rapidities in 
$\{\lambda_1,\dots,\lambda_N\}$;\hfill
\item{} In the remaining set we randomly shift each $\lambda_j$ by
  $\pm \pi/L$ under the constraint that all rapidities in the
  resulting set $\{\mu_1,\dots,\mu_{N-1}\}$ must be different. In this
  way each $\mu_n$ is "paired" with one of the $\lambda_j$ in the
  sense that their difference is as small as possible. 
\end{itemize}
The set $\mathfrak{S}_0(\boldsymbol{\lambda})$ of states
$\langle\boldsymbol{\mu}|$ constructed in this way is clearly
exponentially large in system size. We may characterize
$\mathfrak{S}_0(\boldsymbol{\lambda})$ in terms of a distance 
($\lambda_j=2\pi I_j/L$, $\mu_k=2\pi J_k/L$)
\be
d(\bla,\bmu)\equiv\displaystyle{{\text{min}}_{P\in
    S_N}}\sum_{j=1}^{N-1}(J_j-I_{P_j})^2 \ ,
\ee
as the set of all eigenstates such that  
\be
d(\bla,\bmu)=\frac{N-1}{4}.
\ee
We observe that the matrix elements for this class of states are indeed
much larger than for typical thermal states, \emph{cf.}
Fig.~\ref{fig:state_dep}. As the system size is increased the
probability distribution narrows and shifts towards smaller values
(i.e. large matrix elements). We find that
$P(\mathfrak{M}_{\boldsymbol{\lambda},\boldsymbol{\mu}})$ is well
described by a normal distribution. 

While the matrix elements constructed in this way are large, their
contribution to local correlation functions vanishes in the
thermodynamic limit. This is most easily seen by considering the low
density regime. Here the distance between neighbouring integers in a
thermal state is typically much larger than $1$, which makes it easy
to count states.
\begin{enumerate}
\item{} The number of states in $\mathfrak{S}_0(\boldsymbol{\lambda})$
is $N2^{N-1}$ in the low-density limit. The magnitude of the
corresponding matrix elements with the smooth thermal state can be
estimated as 
\be
\mathfrak{M}_{\bla,\bmu}\approx
-D\ln\big[\frac{4}{\pi^2}\big]+\frac{\ln(L)}{L}-\frac{2\ln(\pi/2)}{L}.
\ee
Hence
\be
\frac{|\langle\boldsymbol{\mu}|\Phi(0)|\boldsymbol{\lambda}\rangle|^2}{\langle\boldsymbol{\lambda}|\boldsymbol{\lambda}\rangle\langle\boldsymbol{\mu}|\boldsymbol{\mu}\rangle}
\propto e^{LD\ln\big[\frac{4}{\pi^2}\big]}\ ,
\ee
while the number of states in $\mathfrak{S}_0(\boldsymbol{\lambda})$
scales exponentially with system size
\be
|\mathfrak{S}_0(\boldsymbol{\lambda})|\propto e^{DL\ln(2)}.
\ee
Concomitantly the contribution of such states to two-point functions
vanishes in the thermodynamic limit as $e^{LD\ln(8/\pi^2)}$.

\item{} We next consider the set of states $\mathfrak{S}_1$ that
differs from $\mathfrak{S}_0$ by adding a single ``soft mode'', by
which we refer to one of the $J_j$ differing from its corresponding
$I_k$ by $\pm(m+\frac{1}{2})$ rather than $\pm\frac{1}{2}$ (where we
keep $m={\cal O}(L^0)$). States in
$\mathfrak{S}_1$ have
\be
d(\bla,\bmu)=\frac{N-1}{4}+\frac{(2m+1)^2-1}{4}.
\ee
The same kind of argument as before now gives
\be
\mathfrak{M}_{\bla,\bmu}\sim
-D\ln\big[\frac{4}{\pi^2}\big]+\frac{\ln(L)}{L}+\frac{2\ln\big(
\frac{2(2m+1)}{\pi}\big)}{L}.
\ee
while the number of states increases to
\be
N(N-1)2^{N-1}.
\ee
This shows that the contribution of such states to two-point functions
again vanishes in the thermodynamic limit.
\item{} The above considerations generalize to a finite number of soft
modes. The rare states of interest therefore involve an extensive
number of soft modes, \emph{cf.}
Refs
\onlinecite{granet2020finite,calabrese2012quantum,bertini2014quantum,schuricht2012dynamics}.
It was shown in Ref.~\cite{granet2021low} how to sum over soft modes
in an arbitrary macro-state at low particle density and obtain an
explicit expression for the single-boson Green's function.
\end{enumerate}

\section{Matrix elements in interacting theories}
\label{sec:MEint}

The matrix elements of local operators between two normalized Bethe
states have been derived in Refs ~\cite{korepin1982calculation,slavnov1989calculation,caux2007one,piroli2015exact}.
In the case of the density operator the square of the matrix element
between two normalized eigenstates
$|\pmb{\lambda}\rangle,|\pmb{\mu}\rangle$ with respective numbers of
Bethe roots $N,N'$ reads 
\begin{widetext}
\begin{equation}
\label{ffdensity}
\begin{aligned}
&\frac{|\langle \pmb{\lambda}|\hat{\rho}(0)|\pmb{\mu}\rangle|^2}{\langle\pmb{\lambda}|\pmb{\lambda}\rangle\langle\pmb{\mu}|\pmb{\mu}\rangle}=\delta_{N,N'}\frac{\left(\sum_{i=1}^N \mu_i-\lambda_i\right)^2}{L^{2N}\mathcal{N}_{\pmb{\lambda}}\mathcal{N}_{\pmb{\mu}}}\frac{\prod_{i\neq j}(\lambda_i-\lambda_j)(\mu_i-\mu_j)}{\prod_{i,j}(\lambda_i-\mu_j)^2}\prod_{i\neq j}\frac{\lambda_i-\lambda_j+ic}{\mu_i-\mu_j+ic}\\
&\times \left|\underset{i,j\neq p}{\det}\left[(V_i^+-V_i^-)\delta_{ij}+i(\mu_i-\lambda_i)\prod_{k\neq i}\frac{\mu_k-\lambda_i}{\lambda_k-\lambda_i}\left(\frac{2c}{(\lambda_i-\lambda_j)^2+c^2} -\frac{2c}{(\lambda_p-\lambda_j)^2+c^2}\right)\right] \right|^2\,.
\end{aligned}
\end{equation}
\end{widetext}
Here $p\in\{1,...,N\}$ can be freely chosen,
\begin{equation}
V_i^\pm=\prod_{k=1}^{N}\frac{\mu_k-\lambda_i\pm ic}{\lambda_k-\lambda_i\pm ic}\,,
\end{equation}
and $\mathcal{N}_{\pmb{\lambda}}$ is given by
\begin{equation}
\label{norm}
\mathcal{N}_{\pmb{\lambda}}={\det}\left[ \delta_{ij} \Big(1+\frac{1}{L}\sum_{k=1}^N \frac{2c}{c^2+\lambda_{i,k}^2}\Big)-\frac{1}{L}\frac{2c}{c^2+\lambda_{i,j}^2}\right]\,.
\end{equation}
When presenting results for the statistic of such matrix elements we
will consider logarithmic expressions like
\be
\mathfrak{M}^{\rho}_{\boldsymbol{\lambda},\boldsymbol{\mu}}=
-\frac{1}{L}\ln\Big[\frac{|\langle\boldsymbol{\mu}|\rho(0)|\boldsymbol{\lambda}\rangle|^2}
{\langle\boldsymbol{\lambda}|\boldsymbol{\lambda}\rangle\langle\boldsymbol{\mu}|\boldsymbol{\mu}\rangle}\Big]\ .
\label{MErho}
\ee
In the following we also use the explicit expressions for the matrix elements
of $g_2$ given in \cite{piroli2015exact}. In the case
$P_{\bla}\neq P_{\bmu}$ the following relation holds
\be
\frac{|\langle \pmb{\lambda}|g_2(0)|\pmb{\mu}\rangle|^2}{\langle\pmb{\lambda}|\pmb{\lambda}\rangle\langle\pmb{\mu}|\pmb{\mu}\rangle}=
\frac{|\langle
  \pmb{\lambda}|\hat{\rho}(0)|\pmb{\mu}\rangle|^2}{\langle\pmb{\lambda}|\pmb{\lambda}\rangle\langle\pmb{\mu}|\pmb{\mu}\rangle}
\left(\frac{{\cal J}_{\bla,\bmu}}{6c(P_{\bla}-P_{\bmu})^2}\right)^2,
\label{g2_vs_rho}
\ee
where
\begin{align}
   {\cal J}_{\bla,\bmu}=&(P_{\bla}-P_{\bmu})^4- 4(P_{\bla}-P_{\bmu})
   (\nu^{(3)}_{\bla}-\nu^{(3)}_{\bmu})\nn
   &+ 3(E_{\bla}-E_{\bmu})^2.
\end{align}
The relation \fr{g2_vs_rho} has important consequences, because by
construction we have
\be
P_{\boldsymbol{\lambda}},E_{\boldsymbol{\lambda}},\nu^{(3)}_{\boldsymbol{\lambda}}
\sim L\ .
\ee
This allows us to conclude that
\be
\mathfrak{M}^{g_2}_{\boldsymbol{\lambda},\boldsymbol{\mu}}=
\mathfrak{M}^{\rho}_{\boldsymbol{\lambda},\boldsymbol{\mu}}+{\cal
  O}\big(\frac{\ln(L)}{L}\big)\ ,
\label{diffrhog2}
\ee
which means that up to finite-size corrections the statistical
properties of $\mathfrak{M}^{g_2}_{\boldsymbol{\lambda},\boldsymbol{\mu}}$
and $\mathfrak{M}^{\rho}_{\boldsymbol{\lambda},\boldsymbol{\mu}}$
should be identical. We verify this by explicit numerical computations
below. 
\subsection{Diagonal matrix elements in interacting theories}
\label{sec:diagonal_int}
\begin{figure}[ht]
\centering
\includegraphics[width=0.45\textwidth]{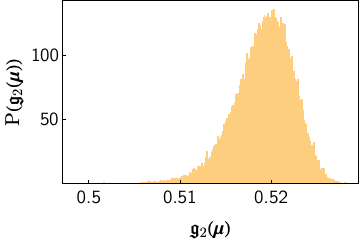}
\caption{Probability distribution of $\mathfrak{g}_2(\bmu)$ for
thermal micro-states at $T=10$, $L=512$ and $c=4$, where
we sample in an energy window as described in the main text.}
\label{fig:diagMEg2}
\end{figure}

\begin{figure}[ht]
    \centering
    \includegraphics[width=0.45\textwidth]{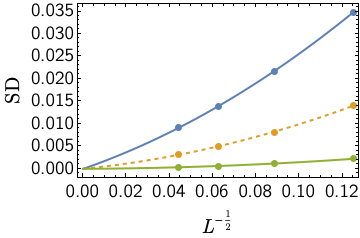}
    \caption{Standard deviations of $P(\mathfrak{g}_2(\bmu))$ for 
thermal micro-states at $T=10$, $D=1$ and at $c=1$ (blue), $c=4$
      (yellow) and $c=16$ (green). The lines are fits as described in
      the main text.}
    \label{fig:diagMEg2SD}
\end{figure}

In order to determine the statistical properties of diagonal matrix
elements for $c<\infty$ we focus on the interaction potential $g_2(x)$
\fr{g2x} because diagonal matrix elements of the density operator
$\rho(x)$ are trivial due to particle number conservation. We
further restrict our analysis to thermal macro-states. We determine
the probability distribution of
\be
\mathfrak{g}_2(\bmu)=\frac{\langle\bmu|g_2(0)|\bmu\rangle}
{\langle\bmu|\bmu\rangle}\ ,
\ee
where $|\bmu\rangle$ are thermal micro-states with energy $E_{\bmu}$,
which we sample in an energy window $\omega=E_{\bmu}-E_{\rm
  smooth}\in[-50,50]$. Here $E_{\rm smooth}$ in the energy of the
smooth thermal micro-state at a given temperature and system size. In
Fig.~\ref{fig:diagMEg2} we show the resulting probability distribution
for $T=10$, $L=512$ and $c=4$.
As we increase the system size the average of the PDF converges as
expected to the thermodynamic limit result. What is of interest is the
scaling of the standard deviation of the PDF with system size. This is
shown for three different values of $c$ and thermal micro-states at
temperature $T=10$ and density $D=1$ in Fig.~\ref{fig:diagMEg2SD} for
system sizes $L=64,128,256,512$. We see that the standard deviation
collapses to zero. Motivated by the results in the $c=\infty$ limit
we have fitted the date to second order polynomials in
$x=L^{-\frac{1}{2}}$ 
\be
f(x)=a_1x + a_2x^2\ .
\ee
The good quality of the fits suggests that for large system sizes the
standard deviation scales as $L^{-1/2}$, as was previously observed
for non-thermal states in the spin-1/2 XXZ chain
\cite{alba2015eigenstate} \footnote{We note however that essentially equally
good descriptions of the data are obtained by two-parameter fits to
$f(x)=a_3 x^{a_4}$.}.
The situation is very different in non-integrable models, where the scaling
of the standard deviation is exponential in system size \cite{steinigeweg2013eigenstate,beugeling2014finitesize,kim2014testing}.

\subsection{Off-diagonal matrix elements in interacting theories}
\label{sec:Offdiagonal}
We now turn to our main topic of interest, off-diagonal matrix
elements in interacting theories. We start by considering matrix
elements of local operators between two different macro-states. On
physical grounds these are expected to be very small and our aim is to
ascertain their scaling with system size at a fixed density.

\subsubsection{Off-diagonal matrix elements between two different thermal macro states}
\label{sec:T1T2int}
\begin{figure}[ht]
\includegraphics[width=0.4\textwidth]{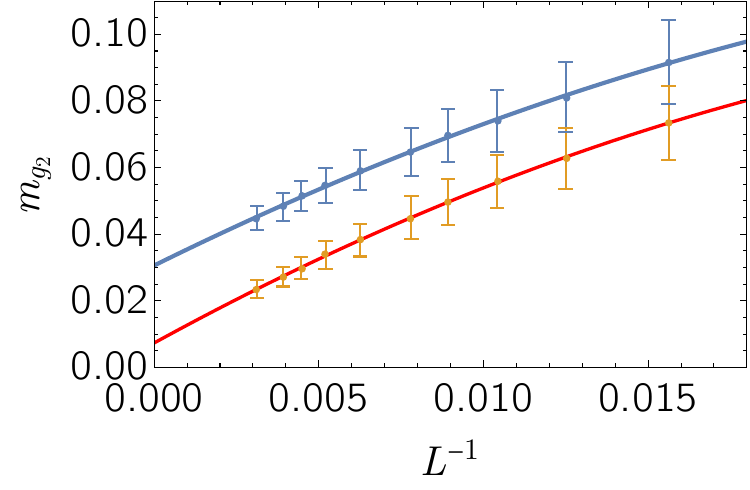}
\caption{Medians (dots) $m_{g_2}$ and standard deviations $s_{g_2}$
(error bars) of the PDFs of $M^{g_2}_{\boldsymbol{\lambda},\boldsymbol{\mu}}/L$ where
$|\boldsymbol{\lambda}\rangle$ and $|\boldsymbol{\mu}\rangle$ are
sampled respectively from thermal macro-states at $T_1=10$, $T_2=5$
(blue symbols) and $T_1=10$, $T_2=7.5$ (red symbols) for $D=1$
in a fixed energy window $|E-Le_{\infty}|<25$.
The solid lines are polynomial fits.}
\label{fig:mg2T1T2}
\end{figure}

\begin{figure}[ht]
\includegraphics[width=0.4\textwidth]{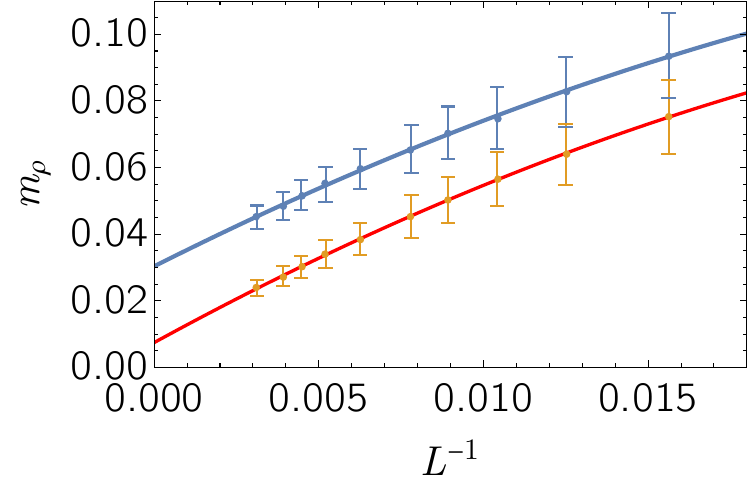}
\caption{Same as Fig.~\ref{fig:mg2T1T2} for matrix elements of the
  density operator $\hat{\rho}$.}
\label{fig:mrT1T2}
\end{figure}

Motivated by our results for matrix elements between two different
macro states in the non-interacting case we examine the
probability distributions of
\begin{align}
\frac{1}{L}\mathfrak{M}^{\cal O}_{\bla,\bmu}&=
\frac{1}{L^2}\ln\left(\frac{|\langle\boldsymbol{\mu}|{\cal
    O}(0)|\boldsymbol{\lambda}\rangle|^2}{\langle\boldsymbol{\lambda}|\boldsymbol{\lambda}\rangle\langle\boldsymbol{\mu}|\boldsymbol{\mu}\rangle}\right)\ ,\quad
     {\cal O}=\hat{\rho},\ g_2.
\label{rho_offdiag}
\end{align}
In particular we focus on the medians $m_{\cal O}$ and standard
deviations $s_{\cal O}$ of the respective PDFs as functions of
particle number $N$, which for simplicity is taken to be equal to $L$
throughout. We generate random samples of $1000$ micro-states
$\langle\boldsymbol{\mu}|$ and $|\boldsymbol{\lambda}\rangle$ that
belong to thermal macro-states at two different temperatures $T_1$ and
$T_2$, and use them to numerically compute $20000$ matrix
elements. Some results from this analysis are shown in
Figs~\ref{fig:mg2T1T2} (for $g_2$) and \ref{fig:mrT1T2} (for
$\hat{\rho}$).   

We observe the following:
\begin{itemize}
\item{} The standard deviations of the PDFs narrow with increasing
  $L$;
\item{} The medians $m_{g_2}$ and $m_\rho$ approach finite limiting
values for large system sizes that depend on the
macro-states but appear to be independent of which of the local
operators $g_2(0)$ and $\hat{\rho}(0)$ we consider.
Numerical extrapolations in $L^{-1}$ give limiting values that are
close to value we obtained for the Bose field in the impenetrable case  
\begin{align}
\frac{1}{L}\mathfrak{M}^{\cal O}_{\bla,\bmu}&\approx
\frac{1}{2}\int_{-\infty}^\infty d\lambda d\mu
\big[\rho_0(\lambda)-\rho_1(\lambda)\big]
\big[\rho_0(\mu)-\rho_1(\mu)\big]\nn
&\qquad\qquad\times\ \ln\big(\lambda-\mu)^2
+o(L^0).
\label{constT1T2}
\end{align}
Here $\rho_0(\lambda)$ and $\rho_1(\lambda)$ are the root densities
of the two macro-states under considerations. The fact that the
extrapolated values for $m_{g_2}$ and $m_{\rho}$ are the same is
easy to understand from the explicit relation \fr{g2_vs_rho} between their
matrix elements, which implies that for $P_{\bla}\neq P_{\bmu}$
\be
\frac{1}{L}\Big[\mathfrak{M}^{g_2}_{\bla,\bmu}-
\mathfrak{M}^{\hat{\rho}}_{\bla,\bmu}\Big]=\frac{1}{L^2}
\ln\left[\frac{{\cal J}_{\bla,\bmu}^2}{6c(P_{\bla}-P_{\bmu})^2}\right]^2.
\ee
The second term of the right-hand-side scales with system size as
$\ln(L)/L^2$ and hence vanishes in the thermodynamic limit.
\end{itemize}

The results of this section are summarized as the following
conjecture: matrix elements involving micro-states belonging to two
different macro-states scale with system size as
\be
\frac{|\langle\boldsymbol{\mu}|{\cal
    O}(0)|\boldsymbol{\lambda}\rangle|^2}{\langle\boldsymbol{\lambda}|\boldsymbol{\lambda}\rangle\langle\boldsymbol{\mu}|\boldsymbol{\mu}\rangle}\propto
e^{-c^{\cal O}_{\rho_0,\rho_1}L^2}\ .
\label{offdiagint}
\ee
Here $c^{\cal O}_{\rho_0,\rho_1}$ is a constant that depends on the
macro-states under consideration and a priori as well on the operator
${\cal O}$. The numerical results presented above are consistent with 
$c^{\cal O}_{\rho_0,\rho_1}$ being independent of ${\cal O}$ and given
by minus the right-hand-side of \fr{constT1T2}. This behaviour is in
stark contrast to the behaviour of off-diagonal matrix elements in
different macro-states non-integrable models as predicted by the ETH.

\subsection{Off-diagonal matrix elements between micro states
  belonging to the same thermal macro states}
\label{sec:T1T1int}
We now turn to matrix elements involving micro-states that belong to
the same macro-state. The question we want to address is how the
corresponding PDFs scale with system size. In order to remove the
sensitive dependence on the ket micro-state we sample the ket states
$|\boldsymbol{\lambda}\rangle$ over the same  energy window as the bra
state $\langle\boldsymbol{\mu}|$. Denoting the energy density in the
thermodynamic limit by $e_\infty$ we take this window to be
$|E-Le_\infty|<25$ in the figures shown below. For simplicity we
focus on thermal macro-states. We observe that we can achieve
excellent data collapse for the PDFs for different system sizes if we
shift the matrix elements by a $L$-dependent constant   
\be
{M}^{\cal O}_{\boldsymbol{\lambda},\boldsymbol{\mu}}=
\mathfrak{M}^{\cal O}_{\boldsymbol{\lambda},\boldsymbol{\mu}}-c^{\cal
  O}_0\ln(L)\ ,\quad
{\cal O}=\hat{\rho},g_2\ .
\label{defMint}
\ee
\begin{figure}[ht]
\includegraphics[width=0.4\textwidth]{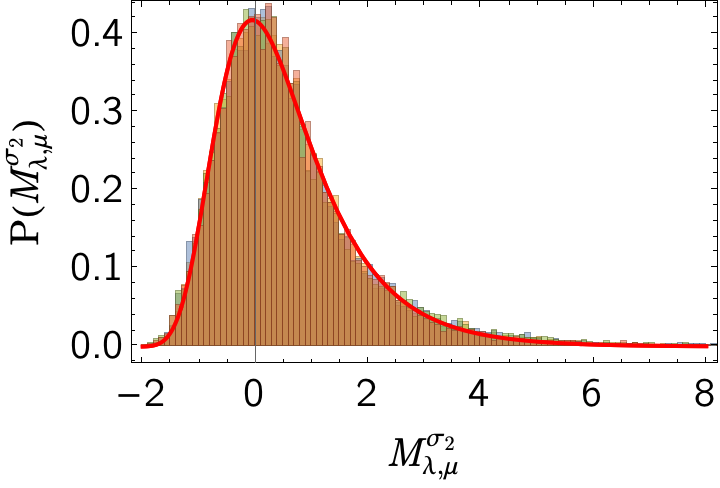}
\caption{Normalized histograms of $M^{g_2}_{\boldsymbol{\lambda},\boldsymbol{\mu}}$ where both  $|\boldsymbol{\lambda}\rangle$ and $|\boldsymbol{\mu}\rangle$ are sampled
from a thermal macro-state at $T=5$, $D=1$ in a fixed energy window $|E-Le_{\infty}|<25$
with $L=N=64,96,160,224$ (respectively in yellow, blue, green and
orange). The solid line is a Fr\'echet distribution function with
fitted parameters $\alpha=16.1378$, $\beta=14.2188$ and $\nu=-14.2223$.}
\label{fig:collapse2_int}
\end{figure}
The resulting probability distributions of appropriately shifted
matrix elements \fr{defM} of the interaction operator $g_2(0)$ in a thermal
macro-state at $T=5$, $D=1$ is shown in Fig.~\ref{fig:collapse2} for
$L=N=64,96,160,224$. 
Here we have fixed the constant in \fr{defM} to be
$c^{\sigma_2}_0=0.754585$, which leads to a very good data
collapse. The resulting probability distribution is well
described by a Fr\'echet distribution with fitted parameters
$\alpha=16.1378$, $\beta=14.2188$ and $\nu=-14.2223$.

The choice $c_0^\rho=c_0^{g_2}$ leads to a good data collapse, as
shown in Fig.~\ref{fig:collapse}, and the resulting PDF is well
described by a Fr\'echet distribution with fitted parameters
$\alpha=17.3467$, $\beta=15.536$ and $\nu=-15.5163$.
\begin{figure}[ht]
\includegraphics[width=0.4\textwidth]{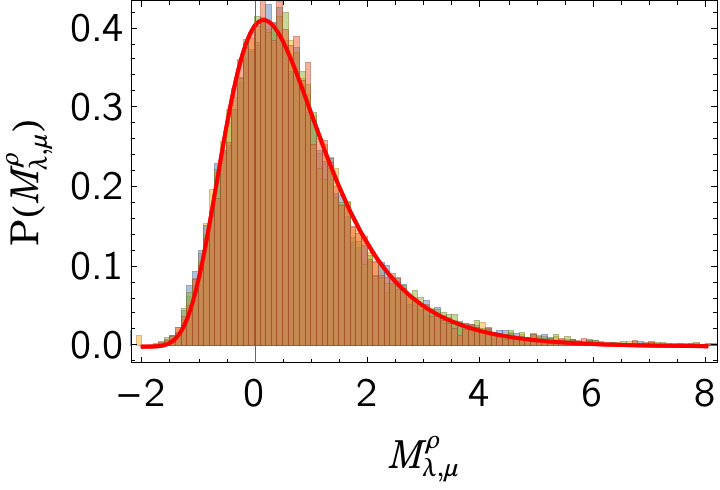}
\caption{Normalized histograms of $M^{\rho}_{\boldsymbol{\lambda},\boldsymbol{\mu}}$ where both  $|\boldsymbol{\lambda}\rangle$ and $|\boldsymbol{\mu}\rangle$ are sampled
from a thermal macro-state at $T=5$, $D=1$ in a fixed energy window $|E-Le_{\infty}|<25$
with $L=N=64,96,160,224$ (respectively in yellow, blue, green and
orange).The solid line is a Fr\'echet distribution function with
fitted parameters $\alpha=17.3467$, $\beta=15.536$ and $\nu=-15.5163$.}
\label{fig:collapse_int}
\end{figure}
The fact that the fitted Fr\'echet distributions differ slightly for
$\hat\rho$ and $g_2$ is a result of the finite-size effects that scale
as $\ln(L)/L$, \emph{cf.} the discussion surrounding \fr{diffrhog2}.

\begin{figure}[ht]
\includegraphics[width=0.4\textwidth]{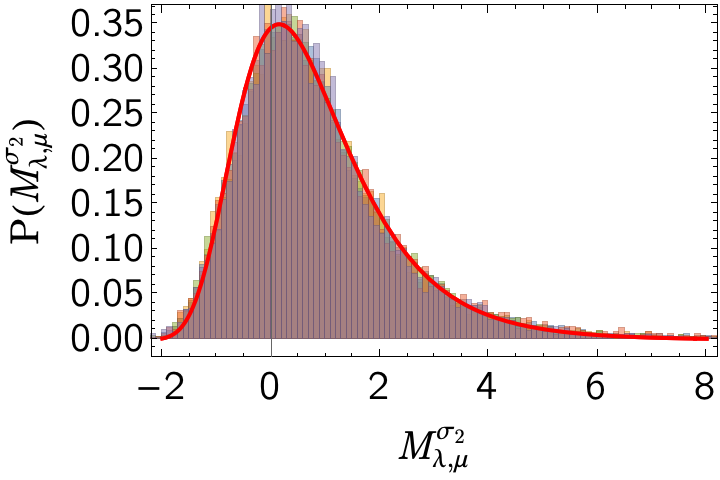}
\caption{Normalized histograms of $M^{g_2}_{\boldsymbol{\lambda},\boldsymbol{\mu}}$ where both  $|\boldsymbol{\lambda}\rangle$ and $|\boldsymbol{\mu}\rangle$ are sampled
from a thermal macro-state at $T=10$, $D=1$ in a fixed energy window $|E-Le_{\infty}|<25$
with $L=N=64,96,128,192,224$ (respectively in yellow, blue, green,
orange and purple). The solid line is a Fr\'echet distribution function with
fitted parameters $\alpha=25.9405$, $\beta=27.2261$ and $\nu=-27.0405$.}
\label{fig:collapse3_int}
\end{figure}

In Fig.~\ref{fig:collapse3_int} we show the probability distributions
of appropriately shifted matrix elements \fr{defM} of the interaction
operator $g_2(0)$ in a thermal macro-state at $T=10$, $D=1$ for
$L=N=64,96,128,192,224$. Our choice of shift parameter
$c^{g_2}=-0.995755$ is again seen to give a good data collapse for the
histograms corresponding to different system sizes, and to be well
described by a fitted Fr\'echet distribution.

The results of this subsection are summarized as the following
conjecture: matrix elements involving micro-states belonging to the same
macro-state scale with system size as
\be
\frac{|\langle\boldsymbol{\mu}|{\cal
    O}(0)|\boldsymbol{\lambda}\rangle|^2}{\langle\boldsymbol{\lambda}|\boldsymbol{\lambda}\rangle\langle\boldsymbol{\mu}|\boldsymbol{\mu}\rangle}\propto
e^{-c^{\cal O}_0L\ln(L)-LM^{\cal O}_{\boldsymbol{\lambda},\boldsymbol{\mu}}}\ ,
\label{offdiagint_same}
\ee
where $c^{\cal O}_0$ depends on the macro-states under consideration
as well as (a priori) on the operator ${\cal O}$. The PDF for
$M^{\cal O}_{\boldsymbol{\lambda},\boldsymbol{\mu}}$, where we sample
both $\boldsymbol{\lambda}$ and $\boldsymbol{\mu}$, is well described
by a Fr\'echet distribution.

\subsection{Atypically large matrix elements}
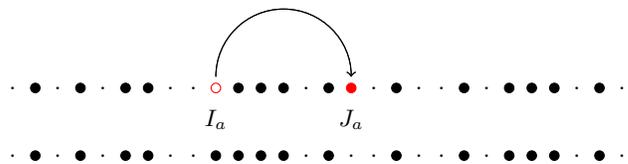
\begin{figure}[ht]
\begin{center}
\begin{tikzpicture}[scale=0.6]
\draw[->]     (3.5,0.25) arc (180: 0:1.5);
\draw[->]     (3.5,0.25) arc (180: 0:1.5);
\node at (-1,0) {.};
\filldraw[black] (-0.5,0) circle (3pt);
\node at (0,0) {.};
\filldraw[black] (0.5,0) circle (3pt);
\node at (1,0) {.};
\filldraw[black] (1.5,0) circle (3pt);
\filldraw[black] (2,0) circle (3pt);
\node at (2.5,0) {.};
\node at (3,0) {.};
\draw[red] (3.5,0) circle (3pt)
node[below=5pt,black] {\small $I_a$};
\filldraw[black] (4,0) circle (3pt);
\filldraw[black] (4.5,0) circle (3pt);
\filldraw[black] (5,0) circle (3pt);
\node at (5.5,0) {.};
\filldraw[black] (6,0) circle (3pt);
\filldraw[red] (6.5,0) circle (3pt)
node[below=5pt,black] {\small $J_a$};
\node at (7,0) {.};
\filldraw[black] (7.5,0) circle (3pt);
\node at (8,0) {.};
\node at (8.5,0) {.};
\filldraw[black] (9,0) circle (3pt);
\node at (9.5,0) {.};
\filldraw[black] (10,0) circle (3pt);
\filldraw[black] (10.5,0) circle (3pt);
\filldraw[black] (11,0) circle (3pt);
\node at (11.5,0) {.};
\filldraw[black] (12,0) circle (3pt);
\node at (12.5,0) {.};
\node at (-1,-1.5) {.};
\filldraw[black] (-0.5,-1.5) circle (3pt);
\node at (0,-1.5) {.};
\filldraw[black] (0.5,-1.5) circle (3pt);
\node at (1,-1.5) {.};
\filldraw[black] (1.5,-1.5) circle (3pt);
\filldraw[black] (2,-1.5) circle (3pt);
\node at (2.5,-1.5) {.};
\node at (3,-1.5) {.};
\filldraw[black] (3.5,-1.5) circle (3pt);
\filldraw[black] (4,-1.5) circle (3pt);
\filldraw[black] (4.5,-1.5) circle (3pt);
\filldraw[black] (5,-1.5) circle (3pt);
\node at (5.5,-1.5) {.};
\filldraw[black] (6,-1.5) circle (3pt);
\node at (6.5,-1.5) {.};
\node at (7,-1.5) {.};
\filldraw[black] (7.5,-1.5) circle (3pt);
\node at (8,-1.5) {.};
\node at (8.5,-1.5) {.};
\filldraw[black] (9,-1.5) circle (3pt);
\node at (9.5,-1.5) {.};
\filldraw[black] (10,-1.5) circle (3pt);
\filldraw[black] (10.5,-1.5) circle (3pt);
\filldraw[black] (11,-1.5) circle (3pt);
\node at (11.5,-1.5) {.};
\filldraw[black] (12,-1.5) circle (3pt);
\node at (12.5,-1.5) {.};
\end{tikzpicture}
\end{center}
\caption{Top: Bethe state that corresponds to a single particle-hole
excitation over the micro-state with half-odd integers shown at the bottom.}
\label{fig:phrho_general}
\end{figure}

As we have seen in the previous section, typical matrix elements of
local operators scale with system size as \fr{offdiagint} or
\fr{offdiagint_same}. Even though there are exponentially many typical
states, they cannot contribute to the Lehmann representation of
two-point functions for the same reasons as discussed below eqn \ref{fxt}.
Hence the matrix elements that matter in spectral representations must be
in the ``nose'' of the PDF and concomitantly be atypically large and
scale exponentially in system size. These involve Bethe states that
differ by a finite number of ``particle-hole'' excitations of their
associated (half-odd) integers. The example of a single particle-hole
excitation is shown in Fig.~\ref{fig:phrho_general}. Given an
eigenstate characterized by the set $\{I_j\}$ (shown as solid circles
at the bottom) we construct an eigenstate characterized by $\{J_j\}$,
obtained by changing a single half-odd integer $I_a$ (red empty
circle) to $J_a$ (red solid circle).  
Carrying out a finite number of particle-hole excitations leads to
matrix elements that are atypically large. The PDF of matrix elements
involving a single particle-hole excitations can be determined
analytically in the large-$c$ limit, as we show next.

\subsubsection{Matrix elements of one-particle-hole states from a \sfix{$1/c$-expansion}}
\label{sssec:1overc}

In Refs~\onlinecite{granet2020a} and \onlinecite{granet2021systematic}
it was shown how to carry out a $1/c$-expansion of 1-particle-hole and
2-particle-hole matrix elements. In order to address statistical
properties of matrix elements we require higher orders in this
expansion. In the following we carry out such an analysis for the
1-particle-hole matrix element of the density operator. Interestingly
this reveals a novel "infrared singularity". We show that these
contributions can be exponentiated, which in turn allows us to get an
explicit expressions for the probability distribution of (the
logarithm of) matrix elements in the thermodynamic limit.

Let $\boldsymbol{\lambda}$ and $\boldsymbol{\mu}$ be solutions to the
Bethe Ansatz equations with corresponding sets of (distinct) integers
$\{I_j\}$ and $\{J_j\}$ respectively, where
\be
J_j=I_j\ ,\quad j\neq a\ ,\quad
J_a=I_a+n.
\ee
This corresponds to a hole with integer $I_a$ (rapidity $\lambda_a$) and a particle with
integer $I_a+n$ (rapidity $\mu_a$).

\begin{figure}[ht]
\begin{center}
\begin{tikzpicture}[scale=0.6]
\draw[->]     (3.5,0.25) arc (180: 0:1.5);
\node at (-1,0) {.};
\filldraw[black] (-0.5,0) circle (3pt);
\node at (0,0) {.};
\filldraw[black] (0.5,0) circle (3pt);
\node at (1,0) {.};
\filldraw[black] (1.5,0) circle (3pt);
\filldraw[black] (2,0) circle (3pt);
\node at (2.5,0) {.};
\node at (3,0) {.};
\draw[red] (3.5,0) circle (3pt)
node[below=5pt,black] {\small $I_a$};
\filldraw[black] (4,0) circle (3pt);
\filldraw[black] (4.5,0) circle (3pt);
\filldraw[black] (5,0) circle (3pt);
\node at (5.5,0) {.};
\filldraw[black] (6,0) circle (3pt);
\filldraw[red] (6.5,0) circle (3pt)
node[below=5pt,black] {\small $I_a+n$};
\node at (7,0) {.};
\filldraw[black] (7.5,0) circle (3pt);
\node at (8,0) {.};
\node at (8.5,0) {.};
\filldraw[black] (9,0) circle (3pt);
\node at (9.5,0) {.};
\filldraw[black] (10,0) circle (3pt);
\filldraw[black] (10.5,0) circle (3pt);
\filldraw[black] (11,0) circle (3pt);
\node at (11.5,0) {.};
\filldraw[black] (12,0) circle (3pt);
\node at (12.5,0) {.};
\end{tikzpicture}
\end{center}
\caption{The set of (half-odd) integers $J_j$ (solid circles)
corresponding to a single particle-hole excitation over a given
microstate characterized by $\{I_j\}$: one (half-odd) integer is
changed from $I_a$ (red empty circle) to $I_a+n$ (red solid circle).} 
\label{fig:phrho}
\end{figure}
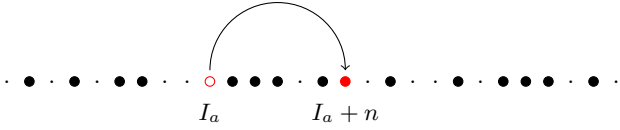
The momentum difference between the two states is
\be
P=\frac{2\pi n}{L}\ .
\ee
In the large-c limit the following expression for the 1-particle-hole
matrix element of the density operator was derived in Ref.~\cite{granet2021systematic}
\begin{widetext}
\begin{align}
\label{ffcm2}
&\frac{|\langle \pmb{\lambda}|\hat{\rho}(0)|\pmb{\mu}\rangle|^2}{\langle\pmb{\lambda}|\pmb{\lambda}\rangle\langle\pmb{\mu}|\pmb{\mu}\rangle}\Bigg|_{\text{1ph}}=
\frac{\alpha^2}{(1+\frac{2}{cL})^2}\frac{1}{L^2}\Bigg[1+\frac{4P'}{cL}\sum_{i\neq a}\Big(\frac{1}{\lambda_{i,a}-P'}-\frac{1}{\lambda_{i,a}}\Big)
+\frac{4P^2}{c^2L^2}\Bigg(-\frac{L^2}{12}\sum_{i\neq a}1+\sum_{\substack{i\neq j\\j\neq a}}\frac{1}{\lambda_{i,j}^2}\nn
&\hskip 5cm+2\Big(\sum_{i\neq a}\frac{1}{\lambda_{i,a}-P'}-\frac{1}{\lambda_{i,a}}\Big)^2
+\sum_{i\neq a}\frac{1}{(\lambda_{i,a}-P')^2}
\Bigg)\Bigg]+{\cal O}(c^{-3})\ .
\end{align}
\end{widetext}
Here we have defined
\be
\lambda_{i,j}=\lambda_i-\lambda_j\ ,\quad \alpha=1+\frac{2D}{c}\ ,\quad P'=\frac{P}{\alpha}\ .
\label{Lprime}
\ee
The leading term is $1/L^2$, but at ${\cal O}(c^{-2})$ there is in
fact an \emph{infrared divergence}
\be
-\frac{DP^2}{3c^2L}
+\frac{4P^2}{c^2L}\left[\frac{1}{L^3}\sum_{\substack{i\neq j\\i,j\neq a}}\frac{1}{\lambda_{i,j}^2}\right] .
\label{sing}
\ee
This contribution scales as $L^{-1}$ whereas the
leading term in the $1/c$-expansion scales as $L^{-2}$. This suggests
the following form for the matrix elements 
\be
\frac{|\langle \pmb{\lambda}|\hat{\rho}(0)|\pmb{\mu}\rangle|^2}{\langle\pmb{\lambda}|\pmb{\lambda}\rangle\langle\pmb{\mu}|\pmb{\mu}\rangle}\bigg|_{\text{1ph}}=
\frac{h_1(\lambda_a,P;\boldsymbol{\lambda})}{L^{2}}e^{-Lf_1(\lambda_a,P;\boldsymbol{\lambda})}\ ,
\label{1phresum}
\ee
where both $h_1$ and $f_1$ have regular expansions in $1/c$
\be
f_1=\frac{a_1}{cL}+\frac{a_2}{c^2}+\ldots\ ,\quad
h_1=1+\frac{b_1}{c}+\dots\ .
\ee
The leading terms in these expansions are then fixed by \fr{ffcm2} and
in particular we have 
\be
a_2=\frac{DP^2}{3}-4P^2\int_{-\infty}^\infty d\mu\  \gamma_{-2}(\mu).
\label{a2gammam2}
\ee
Here the pair distribution function $\gamma_{-2}(\mu)$ is defined as
follows, \emph{cf.} Ref.~\onlinecite{granet2020a}: 
\begin{equation}\label{sipd}
\lim_{L\to\infty}\frac{1}{L^3}\sum_{i\neq j}\frac{g(\lambda_i,\lambda_j)}{(\lambda_i-\lambda_j)^2}
=\int_{-\infty}^\infty d\lambda\
g(\lambda,\lambda)\gamma_{-2}(\lambda)\ ,
\end{equation}
where $g(\lambda,\mu)$ is any smooth function. Importantly this quantity depends on
details of the state $|\boldsymbol{\lambda}\rangle$ beyond its root
density, namely the joint PDF of pairs of Bethe roots. The simplest
way of determining it is by reverting to the sum in \fr{sing}. The
fact that expression for $a_2$ \fr{a2gammam2}  involves the pair
distribution function shows explicitly that the matrix elements
\fr{1phresum} depend on details of the micro-state
$|\boldsymbol{\lambda}\rangle$ beyond the macro-state information
encoded in the particle root density. 

In order to exhibit the structure \fr{1phresum} more fully we have
determined the square of the 1-particle-hole matrix elements up to ${\cal
  O}(c^{-4})$ in Appendix \ref{app:1overc}. Exponentiating the
resulting infrared divergences using the conjecture \fr{1phresum} 
results in 
\begin{widetext}
\begin{align}
\label{ff1phres}
&\frac{|\langle \pmb{\lambda}|\hat{\rho}(0)|\pmb{\mu}\rangle|^2}{\langle\pmb{\lambda}|\pmb{\lambda}\rangle\langle\pmb{\mu}|\pmb{\mu}\rangle}\Bigg|_{\text{1ph,res}}
\approx\frac{\alpha^2}{L^2\big(1+\frac{2}{cL}\big)^2}
\Bigg[1-\frac{4P'}{cL}\big(\Gamma_{-1}-\tilde{\Gamma}_{-1}\big)
+\frac{8(P')^2}{c^2L^2}\big(\Gamma_{-1}-\tilde{\Gamma}_{-1}\big)^2
+\frac{4(P')^2}{c^2L^2}\big(\Gamma_{-2}+\tilde{\Gamma}_{-2}\big)\nn
&\qquad-\frac{4\Gamma_2}{c^3L}
-\frac{12P^2}{c^3L}+\frac{4P^3}{3c^3L}(\Gamma_{-1}-\tilde{\Gamma}_{-1})
-\frac{16P^3}{3c^3L^3}\Big[2\big(\Gamma_{-1}-\tilde{\Gamma}_{-1}\big)^3+3\big(\Gamma_{-1}-\tilde{\Gamma}_{-1}\big) \big(\Gamma_{-2}+\tilde{\Gamma}_{-2}\big)+\big(\Gamma_{-3}-\tilde{\Gamma}_{-3}\big)\Bigg]\nn
&\qquad\times\exp\Big[-\frac{P^2(N-1)}{3c^2}+\frac{4(P')^2}{c^2L^2}\sum_{\substack{i\neq j\\i,j\neq a}}\frac{1}{\lambda_{ij}^2}+\frac{4P^2D(N-1)}{c^3}-\frac{4PD\Gamma_1}{c^3}
-\frac{8}{c^3L}\Big(\Gamma_1^2-N\Gamma_2\Big)\Big].
\end{align}
\end{widetext}
Here we have introduced shorthand notations
\be
\Gamma_n=\sum_{i\neq a}(\lambda_{i,a})^n\ ,\
\tilde{\Gamma}_n=\sum_{i\neq a}(\lambda_{i,a}-P')^n\ .
\ee
We have verified numerically that the expression \fr{ff1phres}
provides a good approximation to the exact matrix element at large
values of $c$ in the regime $L>c$. This supports the conjecture
\fr{1phresum}. Some of this evidence is presented in Appendix
\ref{app:1overc}.

Comparing \fr{ff1phres} with \fr{1phresum} and dropping terms that
vanish in the thermodynamic limit we have 
\begin{align}
f_1(\lambda_a,P;\boldsymbol{\lambda})&=
\frac{P^2D}{3c^2}-\frac{4{P'}^2}{c^2L^3}\sum_{i\neq
j}\frac{1}{\lambda_{ij}^2}
-\frac{4P^2D^2}{c^3}\nn
&+\frac{4PD\Gamma_1}{Lc^3}
+\frac{8}{c^3L^2}\Big(\Gamma_1^2-N\Gamma_2\Big)+\dots
\end{align}
In order to consider asymptotically large systems it is useful to
express $f_1$ in terms of the particle and hole rapidities using
\be
\lambda_p=\lambda_h+P'\big(1+\frac{2}{cL}\big)+{\cal O}(c^{-3}).
\ee
Assuming for simplicity that the root distribution function
$\rho(\lambda)$ of the rapidities $\bla$ is an even function from here
on we have
\begin{align}
f_1(\lambda_h,\lambda_p;\boldsymbol{\lambda})&=
\frac{P^2}{c^2}\Big[\frac{D}{3}-\frac{4D^2}{c}-\frac{4}{\alpha^2}\int
d\mu\ \gamma_{-2}(\mu)\Big]\nn
&-\frac{4D^2}{c^3}\lambda_h(\lambda_p-\lambda_h)-\frac{8D}{c^3}\int
d\mu\ \mu^2\ \rho(\mu)\nn
&+o(L^0)\ .
\end{align}
Here we retain the label $\boldsymbol{\lambda}$ in order to indicate
that $f_1$ depends on properties of the ket $|\bla\rangle$ beyond
those encoded in its root density $\rho(\lambda)$. In the large-$L$
limit the logarithm of the matrix elements \fr{MErho} for our one
particle-hole excitation then becomes
\be
\mathfrak{M}^{\rho}_{\boldsymbol{\lambda},\boldsymbol{\mu}}=f_1(\lambda_h,\lambda_p;\bla)+o(L^0)\ .
\ee
We now can determine the probability distribution $P(z,Q)$ of
$\mathfrak{M}^{\rho}_{\boldsymbol{\lambda},\boldsymbol{\mu}}\Big|_{\rm
  1ph}$ at a fixed momentum transfer $Q$ between the two states. To
that end we introduce a rapidity cutoff $\Lambda$ that translates into
a cutoff $I^{(L)}_{\rm max}$ for our Bethe integers,
i.e. we consider only solutions of the Bethe equations such that all
(half-odd) integers fulfil $|I_j|<I^{(L)}_{\rm max}$. We then define 
\be
P_L^\eta(z,Q)=\frac{1}{N_1}\!\sum_{I_p,I_h}\delta_\eta(z-
\mathfrak{M}^{\rho}_{\boldsymbol{\lambda},\boldsymbol{\mu}
})\ \delta_\eta\big(Q-\frac{2\pi(I_p-I_h)}{L}\big),
\ee
where $N_1$ is the total number of one particle-hole
excitations given the state $|\bla\rangle$ and $I^{(L)}_{\rm max}$. We are
interested in the joint PDF
\be
{\rm P}(z,Q)=\lim_{\eta\to0}\lim_{L\to\infty}P_L^\eta(z,Q).
\ee
Turning sums into integrals in the thermodynamic limit gives
\begin{align}
&{\rm P}(z,Q)=\frac{1}{{\cal N}_\Lambda}\int_{-\Lambda}^\Lambda
d\lambda\ \rho(\lambda)\int_{-\Lambda}^{\Lambda} d{\mu}\ \rho_h(\mu)\nn
&\times \delta\big(z-f_1(\lambda,\mu;\boldsymbol{\lambda})\big)\
\delta\big(Q-2\pi[z(\mu)+z(\lambda)]\big)\ ,
\end{align}
where $z(\lambda)$ is the counting function defined in \fr{count} and
\be
{\cal N}_\Lambda=\int_{-\Lambda}^\Lambda
d\lambda\ \rho(\lambda)\int_{-\Lambda}^{\Lambda} d{\mu}\ \rho_h(\mu)\ .
\ee
In order to proceed we now use the approximation
\be
2\pi[z(\mu)-z(\lambda)]=(\mu-\lambda)\alpha+{\cal O}(c^{-3}).
\ee
This allows us to carry out the integral over $\mu$ in an
elementary fashion
\be
{\rm P}(z,Q)\approx\frac{1}{\alpha{\cal N}_\Lambda}\sum_n\frac{\rho_p(x_n)\ \rho_h(x_n+Q/\alpha)}
{|f'_1(x_n,x_n+Q/\alpha)|}\ ,
\ee
where the sum is over solutions $|x_n|<\Lambda$ to the equation
\be
z-f_1(x_n,x_n+Q/\alpha;\boldsymbol{\lambda})=0.
\ee
In our case there is only a single solution 
\begin{align}
x_1(z,Q) & =-\frac{zc^3\alpha}{4D^2Q}- \frac{2\alpha}{DQ}\int d\mu\ \mu^2\rho(\mu)\nn
&+\frac{Qc}{4D^2\alpha}\Big[\frac{D}{3}-\frac{4D^2}{c}-\frac{4}{\alpha^2}\int
d\mu \gamma_{-2}(\mu)\Big],
\end{align}
so that we arrive at a very simple answer
\be
{\rm P}(z,Q)\approx\frac{c^3}{4D^2{\cal N}_\Lambda}\frac{\rho(x_1)\ \rho_h(x_1+Q/\alpha)}
{|Q|}\ .
\label{TDlimit}
\ee

\begin{figure}[ht]
\includegraphics[scale=0.5]{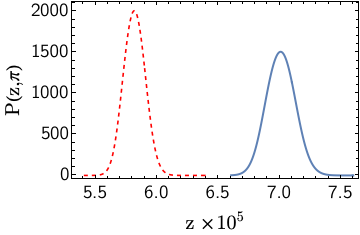}
\caption{Probability distribution functions $P(z,Q=\pi)$ for smooth micro-states corresponding to thermal macro-states with $D=0.25$, $\beta=0.25$, $c=100$ (blue) and $c=110$ (red dashed).}
\label{fig:probdist1}
\end{figure}

The probability distribution functions $P(z,Q=\pi)$ for smooth
micro-states corresponding to a thermal macro-state with $D=0.25$,
$\beta=0.25$ and two values of $c$ are shown in
Fig.~\ref{fig:probdist1}. 
We see that as expected the probability distribution function narrows and is peaked at smaller values of $z$ as $c$ is increased. In the limit $c\to\infty$ we know that $f_1\to 0$.

Here we have computed $\int d\mu\ \gamma_{-2}(\mu)$ by considering the
scaling of its finite-size expression with $L$ for large systems of up
to $L=4096$. It is important to stress that the thermodynamic limit
result shown in Fig.~\ref{fig:probdist1} is approached only for very large
$L$ such that $Lf_1\gg 1$. The results of this subsection are
summarized as follows:
\begin{itemize}
\item{} Matrix elements involving a single-particle hole excitation over a
given micro-state at a finite energy density relative to the ground
state are exponentially small in system size
  \be
\frac{|\langle \pmb{\lambda}|\hat{\rho}(0)|\pmb{\mu}\rangle|^2}{\langle\pmb{\lambda}|\pmb{\lambda}\rangle\langle\pmb{\mu}|\pmb{\mu}\rangle}\bigg|_{\text{1ph}}=
\frac{h_1(\lambda_a,P;\boldsymbol{\lambda})}{L^{2}}e^{-Lf_1(\lambda_a,P;\boldsymbol{\lambda})}\ ,
\ee
where the function $f_1$ has depends on details of the micro-state
beyond the root density $\rho(\lambda)$ that specifies the macro-state
to which it belongs. Above we have derived explicit expressions for
the $1/c$-expansions of the functions $h_1$ and $f_1$. 
\item{} These matrix elements are very large compared to typical
matrix elements in the same macro-state, which scale as \fr{offdiagint_same}.
\item{} If we introduce a momentum cutoff there are only polynomially
many (in system size) single particle-hole excitations, which means
that they do not contribute to the thermodynamic limit of Lehmann
representations of two-point functions for the same reasons as
discussed below eqn \ref{fxt}. 
\end{itemize}

\subsubsection{Multiple particle-hole excitations}
\label{sssec:multipleph}
We have verified numerically that making a fixed number of
particle-hole excitations over a micro-state belonging to a thermal
macro-state again leads to matrix elements that are exponentially small in
system size. Hence such matrix elements are also anomalously large
compared to typical ones. However, given a momentum cutoff there are
only polynomially many (in system size) $m$-particle-hole
excitations (where $m$ is fixed), which means that they do not
contribute to the thermodynamic limit of Lehmann representations of
two-point functions for the same reasons as discussed below eqn \ref{fxt}. 
The states that do contribute to such Lehmann representations involve
an extensive number of particle-hole excitations.
\section{Summary and conclusions}
\label{sec:Conclusions}
In this work we have used integrability methods to determine the
structure of matrix elements of local operators in energy eigenstates
of the Lieb-Liniger model. The latter is a key paradigm of integrable
many-particle quantum systems and has the distinctive property that
energy eigenstates at an arbitrary energy density can be understood in
terms of a single species of elementary excitations, which greatly
simplifies the task of numerically determining matrix elements for
large system sizes. The existence of an extensive number of mutually
compatible conserved charges affects the structure of energy
eigenstates at finite energy densities: in addition to thermal states
we can have other macro-states that differ in their values of the
densities of some of the conserved charges. Our results for the
structure of matrix elements of local operators in the Lieb-Liniger
model can be summarized as follows.

\begin{itemize}
\item{} Typical diagonal matrix elements in a macro-state characterized by a
root density $\rho(\lambda)$ depend only on macro-state information up
to finite-size corrections
\be
\frac{\langle\bmu|{\cal O}(0)|\bmu\rangle}{\langle\bmu|\bmu\rangle}=
f_{\cal O}(\rho)+{\cal O}(L^{-\frac{1}{2}})\ .
\ee
Here $f_{\cal O}(\rho)$ is a function that depends smoothly on the
densities of the conserved charges. This can be thought of as a
natural generalization of the eigenstate thermalization hypothesis for
diagonal matrix elements. Our findings are in agreement with previous
work on diagonal matrix elements in non-thermal states in the the
spin-1/2 XXZ chain\cite{alba2015eigenstate}.
\item{} Typical off-diagonal matrix elements involving micro-states belonging to two
different macro-states scale with system size as
\be
\frac{|\langle\boldsymbol{\mu}|{\cal
    O}(0)|\boldsymbol{\lambda}\rangle|^2}{\langle\boldsymbol{\lambda}|\boldsymbol{\lambda}\rangle\langle\boldsymbol{\mu}|\boldsymbol{\mu}\rangle}\propto
e^{-c^{\cal O}_{\rho_0,\rho_1}L^2}\ .
\ee
Here $c^{\cal O}_{\rho_0,\rho_1}$ is a constant that depends on the
macro-states under consideration and a priori as well on the operator
${\cal O}$.
\item{} Typical off-diagonal matrix elements involving micro-states belonging to the same
macro-state scale with system size as
\be
\frac{|\langle\boldsymbol{\mu}|{\cal
    O}(0)|\boldsymbol{\lambda}\rangle|^2}{\langle\boldsymbol{\lambda}|\boldsymbol{\lambda}\rangle\langle\boldsymbol{\mu}|\boldsymbol{\mu}\rangle}\propto
e^{-c^{\cal O}_0L\ln(L)-LM^{\cal O}_{\boldsymbol{\lambda},\boldsymbol{\mu}}}\ ,
\label{summary_same}
\ee
where $c^{\cal O}_0$ depends on the macro-states under consideration
as well as (a priori) on the operator ${\cal O}$. The PDF for
$M^{\cal O}_{\boldsymbol{\lambda},\boldsymbol{\mu}}$, where we sample
both $\boldsymbol{\lambda}$ and $\boldsymbol{\mu}$, is well described
by a Fr\'echet distribution. If we fix the ket state
$|\boldsymbol{\lambda}\rangle$ and consider the PDF of matrix elements
obtained by sampling $|\boldsymbol{\mu}\rangle$ we observe a strong
dependence on the details of $|\boldsymbol{\lambda}\rangle$, i.e. the
multivariate probability distribution of half-odd
integers. Nevertheless, if we fix $|\boldsymbol{\lambda}\rangle$ to be
the smooth thermal state the PDF for $M^{\cal
  O}_{\boldsymbol{\lambda},\boldsymbol{\mu}}$ is well characterized by
a Fr\'echet distribution (with different parameters compared to the
case where we sample both $\boldsymbol{\lambda}$ and
$\boldsymbol{\mu}$).
\item{} There are rare, but still exponentially many (in particle
number, given a momentum cut-off), matrix elements between eigenstates
that belong to the same macro-state that are much larger than
\fr{summary_same}, but instead are merely exponentially small in
system size
\be
\frac{|\langle\boldsymbol{\mu}|{\cal
    O}(0)|\boldsymbol{\lambda}\rangle|^2}{\langle\boldsymbol{\lambda}|\boldsymbol{\lambda}\rangle\langle\boldsymbol{\mu}|\boldsymbol{\mu}\rangle}\bigg|_{\rm
  rare}\propto
e^{-LM^{\cal  O}_{\boldsymbol{\lambda},\boldsymbol{\mu}}}\ .
\ee
These can be characterized by the property that the sets of (half-odd)
integers corresponding to the Bethe roots $\boldsymbol{\lambda}$ and
$\boldsymbol{\mu}$ are atypically close to one another. For the case
of the density operator we obtained explicit results for the simplest
such matrix elements by generalizing the $1/c$-expansion method
pioneered in Refs~\cite{granet2020a,granet2021systematic}.
\item{} The observed structure of off-diagonal matrix elements in
interacting theories is very similar to the one we find for the Bose
field in the $c\to\infty$ limit, in which the Lieb-Liniger model
can be mapped to free fermions. The origin of this similarity is the
structure of the singularities of the matrix elements when viewed as
functions of the spectral parameters (in a large finite volume the
fact that the spectral parameters must fulfil the Bethe equations
regularizes these singularities). 
\end{itemize}
Our work poses a number of important questions that should be
addressed in future work. First and foremost, it should be clarified
whether the results obtained here indeed carry over to all local
operators in the Lieb-Liniger model, and to other integrable models,
as we conjecture. To that end it would be useful to consider
non-thermal macro-states in the spin-1/2 XXZ chain as was 
done in \cite{alba2015eigenstate} and check whether the same kind of
scaling behaviour of matrix elements with system size reported here
occurs. An analysis of the matrix elements of the Bose field in the
Lieb-Liniger model for $0<c<\infty$ will be reported elsewhere
\cite{wip}.                     
Second, one should attempt to conduct an analogous study in
models that feature bound states (string solutions to the Bethe
equations). This appears difficult at present and will require a
better control of matrix elements involving strings than is 
available in the literature. Third, the statistical properties of the
rare, large matrix elements should be investigated in more detail
in the case where one has a finite but low density of particle-hole
excitations. Here the hope would be to find a way to randomly sample
the large matrix elements that dominate the Lehmann representations of
two-point functions and related quantities of interest
\cite{bulchandani2022onset}. 

\begin{acknowledgments}
We are very grateful to J.-S. Caux and N. Robinson for collaborating
with us during the early stages of this work and numerous very helpful
discussions and suggestions. This work was supported by the EPSRC under grant
EP/S020527/1 (FHLE) and the European Research Council under ERC
Advanced grant No 743032 DYNAMINT (AJJMdK).
\end{acknowledgments}
Author contributions: FHLE conceptualized the work, carried out all
calculations and computations and wrote the manuscript. 
AJJMdK worked on carrying out numerical calculations for the
interacting case and analyzing the associated results. 
\appendix
\section{\sfix{$1/c$}-expansion of the 1-particle-hole matrix element}
\label{app:1overc}
In this Appendix we present some details regarding the $1/c$-expansion
of the density matrix element \fr{MErho} between two Bethe states
differing by a single particle-hole excitation, \emph{cf.} section
\ref{sssec:1overc} of the main text. For convenience we first recall
some of the notations introduced in the main text. We consider two solutions
$\boldsymbol{\lambda}$ and $\boldsymbol{\mu}$ to the Bethe Ansatz
equations with corresponding sets of (distinct) integers $\{I_j\}$ and
$\{J_j\}$, where 
\be
J_j=I_j\ ,\quad j\neq a\ ,\quad
J_a=I_a+n.
\ee
This corresponds to a hole with integer $I_a$ (rapidity $\lambda_a$) and a particle with
integer $I_a+n$ (rapidity $\mu_a$). The momentum difference between
the two states is $P=2\pi n/L$. In order to simply the expression for
the matrix element in the $1/c$-expansion it is useful to introduce
short-hand notations 
\begin{align}
\lambda_{i,j}&=\lambda_i-\lambda_j\ ,\quad
\alpha=1+\frac{2D}{c}\ ,\quad P'=\frac{P}{\alpha}\ ,\nn
\Gamma_n&=\sum_{i\neq a}(\lambda_{i,a})^n\ ,\
\tilde{\Gamma}_n=\sum_{i\neq a}(\lambda_{i,a}-P')^n\ .
\end{align}
Solving the Bethe equations in the framework of the $1/c$-expansion gives
\begin{align}
\label{besolve}
\lambda_i=&\frac{2\pi I_i}{\alpha L}+\frac{4\pi
}{cL^2\alpha}\sum_{j=1}^N
I_j+\frac{(2\pi)^4}{3\pi
  c^3(\alpha L)^4}\sum_{j=1}^N
\left({I_i-I_j}\right)^3\nn
&+{\cal O}(c^{-4})\,.
\end{align}
We then can express the rapidities $\mu_k$ in terms of the $\lambda_j$ as
\begin{align}
\mu_i=&\lambda_i+\frac{2P'}{cL}\left[1-\frac{(\lambda_{i,a})^2}{c^2}+\frac{P\lambda_{i,a}}{c^2}-\frac{P^2}{3c^2}\right]+{\cal
  O}(c^{-4})\ ,\nn
\mu_a=&\lambda_a+P'\left[1+\frac{2}{cL}\right]+\frac{2}{3c^3L}\sum_{k\neq
  a}\left[(\lambda_{a,k}+P)^3-\lambda_{a,k}^3\right]\nn
&+   {\cal O}(c^{-4})\ .
\end{align}
We now extend the analysis of Ref.~\cite{granet2020a} by carrying out
a $1/c$-expansion of the various factors in the expression
\fr{ffdensity} of the matrix elements of th density operator up to order
${\cal O}(c^{-4})$. As in \cite{granet2020a} we retain certain
contributions to all orders. We find
\begin{align}
&A_1\equiv\prod_{i\neq
j}\frac{\lambda_{i,j}+ic}{\mu_{i,j}+ic}\simeq 1-\frac{{P'}^2(N-1)}{c^2}+\frac{2P'}{c^2}\Gamma_1(\lambda_a),\nn
&A_2\equiv\Big|\prod_{i\neq a}(V_i^+-V_i^-)\Big|^2\simeq\Big(\frac{2P}{c}\Big)^{2N-2}\Big(1-\frac{2}{c^2}\Gamma_2(\lambda_a)\Big),\nn
&A_3\equiv\prod_{\substack{i\neq j\\ i\neq a\\ j\neq
    a}}\frac{\lambda_{i,j}\mu_{i,j}}{(\lambda_i-\mu_j)^2}\simeq1+\frac{4{P'}^2}{c^2L^2}
\sum_{\substack{i\neq j\\ i\neq a\\ j\neq  a}}\frac{1}{\lambda_{i,j}^2},\nn
&\mathcal{N}_{\pmb{\lambda}}\simeq\alpha^{N-1}\Big[1+\frac{4}{c^3
L}\Big(\big(\sum_i\lambda_i\big)^2-N\sum_i\lambda_i^2\Big)\Big],\nn
&\mathcal{N}_{\pmb{\mu}}\simeq\mathcal{N}_{\pmb{\lambda}}\Big[1+\frac{4P}{c^3L}\big(2\Gamma_1(\lambda_a)-(N-1)P\big)\Big],
\label{as1}
\end{align}
\begin{align}
&A_4\equiv\prod_{i\neq a}\frac{\lambda_{i,a}^2}{(\mu_i-\lambda_a)^2}\simeq
1-\frac{4P'}{cL}\Gamma_{-1}+\frac{8{P'}^2}{c^2L^2}\Big[\Gamma_{-1}^2+\frac{\Gamma_{-2}}{2}\Big]\nn
&\hskip2.2cm-\frac{16P^3}{3c^3L^3}\Big[2\Gamma_{-1}^3+3\Gamma_{-1}\Gamma_{-2}+\Gamma_{-3}\Big]\nn
&\hskip2.2cm +\frac{4P}{c^3L}\Big[\Gamma_{1}-3P+\frac{P^2\Gamma_{-1}}{3}\Big].
\end{align}
\begin{align}
&A_5\equiv\prod_{i\neq a}\frac{(\mu_i-\mu_a)^2}{(\lambda_i-\mu_a)^2}\simeq
1+\frac{4P'}{cL}\big(1-\frac{P^2}{3c^2}\big)\tilde{\Gamma}_{-1}\nn
&\hskip 1.7cm+\frac{4{P'}^2}{c^2L^2}\Big[2\tilde{\Gamma}_{-1}^2+\tilde{\Gamma}_{-2}\Big]-\frac{4P}{Lc^3}\Gamma_1\nn
&\hskip1.7cm+\frac{16P^3}{3c^3L^3}\Big[2\tilde{\Gamma}_{-1}^3+3\tilde{\Gamma}_{-2}\tilde{\Gamma}_{-1}+\tilde{\Gamma}_{-3})\Big],
\end{align}
\begin{align}
&A_6\equiv\prod_{i=1}^N\frac{1}{(\lambda_i-\mu_i)^2}\simeq\frac{\alpha^{2N}}{P^2}\left(\frac{Lc}{2P}\right)^{2N-2}\nn
&\times\Big[
1+\frac{2}{c^2}\big(\Gamma_2-P\Gamma_1+\frac{P^2}{3}(N-1)\big)\big(1-\frac{2}{cL}\big)\Big],
\label{as2}
\end{align}
Finally we have
\begin{align}
&i(\mu_l-\lambda_l)\prod_{k\neq
l}\frac{\mu_k-\lambda_l}{\lambda_k-\lambda_l}\Big[\frac{2c}{\lambda_{l,j}^2+c^2}
-\frac{2c}{(\lambda_{p,j}^2+c^2}\Big]\simeq 0,\nn
&\Big(\sum_i\mu_i-\lambda_i\Big)^2\simeq P^2\ .
\label{as3}
\end{align}
The leading corrections in \fr{as1}-\fr{as3} are of order ${\cal
  O}(c^{-4})$. Substituting the results back into the expression
\fr{MErho} for the matrix elements leads to the following expression
for one states that differ by a single particle-hole excitation
\begin{widetext}
\begin{align}
\label{ff1ph3}
&\frac{|\langle \pmb{\lambda}|\hat{\rho}(0)|\pmb{\mu}\rangle|^2}{\langle\pmb{\lambda}|\pmb{\lambda}\rangle\langle\pmb{\mu}|\pmb{\mu}\rangle}\bigg|_{\text{1ph}}=\frac{\alpha^2}{L^2\big(1+\frac{2}{cL}\big)^2}
\Bigg[1-\frac{4P'}{cL}\big(\Gamma_{-1}-\tilde{\Gamma}_{-1}\big)-\frac{P^2(N-1)}{3c^2}+\frac{4{P'}^2}{c^2L^2}\sum_{i\neq
j}\frac{1}{\lambda_{ij}^2}
+\frac{8{P'}^2}{c^2L^2}\big(\Gamma_{-1}-\tilde{\Gamma}_{-1}\big)^2\nn
&+\frac{4{P'}^2}{c^2L^2}\big(\Gamma_{-2}+\tilde{\Gamma}_{-2}\big)
+\frac{4P^3N}{3c^3L}\big(\Gamma_{-1}-\tilde{\Gamma}_{-1}\big)
+\frac{4P^2D(N-1)}{c^3}-\frac{4PD\Gamma_1}{c^3}-\frac{4\Gamma_2}{c^3L}
-\frac{8}{c^3L}\Big(\big(\sum_i\lambda_i\big)^2-N\sum_i\lambda_i^2\Big)-\frac{12P^2}{c^3L}\nn
&-\frac{16P^3}{3c^3L^3}\Big[2\big(\Gamma_{-1}-\tilde{\Gamma}_{-1}\big)^3+3\big(\Gamma_{-1}-\tilde{\Gamma}_{-1}\big) \big(\Gamma_{-2}+\tilde{\Gamma}_{-2}\big)+\big(\Gamma_{-3}-\tilde{\Gamma}_{-3}\big)\Big]
-\frac{16P^3}{c^3L^3}\big(\Gamma_{-1}-\tilde{\Gamma}_{-1}\big)
\sum_{i\neq j}\frac{1}{\lambda_{ij}^2}\Bigg]+{\cal O}(c^{-4})\ .
\end{align}
\end{widetext}
This expression indeed exhibits ``infrared divergences'',
i.e. contributions that acquire additional factors of $L$ compared to
the leading term, that are compatible with the conjecture
\fr{1phresum}. We conjecture that these terms can be exponentiated and
in order to do so it is useful to return to the individual factors
they arise from 
\begin{align}
A_1&\approx A_1^\text{(r)}=
\exp\Big(-\frac{{P'}^2(N-1)}{c^2}+\frac{2P'}{c^2}\Gamma_1\Big)\ ,\nn
A_2&\approx A_2^\text{(r)}=\Big(\frac{2P}{c}\Big)^{2N-2}\exp\Big(-\frac{2}{c^2}\Gamma_2\Big) ,\nn
A_3&\approx A_3^\text{(r)}=\exp\bigg(\frac{4{P'}^2}{c^2L^2}
\sum_{\substack{i\neq j\\ i,j\neq a}}\frac{1}{\lambda_{i,j}^2}\bigg)\ ,\nn
\mathcal{N}_{\pmb{\lambda}}&\approx\mathcal{N}^\text{(r)}_{\pmb{\lambda}}
=\alpha^{N-1}\exp\Big(\frac{4}{c^3
L}\big(\Gamma_1^2-N\Gamma_2\big)\Big)\ ,
\label{resum1}
\end{align}
\begin{align}
A_6&\approx A_6^\text{(r)}=\frac{\alpha^{2N}}{P^2\big(1+\frac{2}{cL}\big)^2}\left(\frac{Lc}{2P}\right)^{2N-2}\nn
&\times\exp\Big(
\frac{2}{c^2}\big(\Gamma_2-P\Gamma_1+\frac{P^2}{3}(N-1)\big)\big(1-\frac{2}{cL}\big)\Big).
\label{resum2}
\end{align}
Using \fr{resum1}, \fr{resum2} to exponentiate the infrared
singularities in \fr{ff1ph3} results in the expression \fr{ff1phres}
in the main text. In order to assess the accuracy of \fr{ff1phres} we
have computed its ratio $R$ to numerically exact matrix elements for a number of
particle-hole excitations over a "smooth" thermal state with
$\beta=1$, where
\be
R=
\frac{|\langle \pmb{\lambda}|\hat{\rho}(0)|\pmb{\mu}\rangle|^2}{\langle\pmb{\lambda}|\pmb{\lambda}\rangle\langle\pmb{\mu}|\pmb{\mu}\rangle}\bigg|_{\text{1ph,res}}
\bigg(\frac{|\langle \pmb{\lambda}|\hat{\rho}(0)|\pmb{\mu}\rangle|^2}{\langle\pmb{\lambda}|\pmb{\lambda}\rangle\langle\pmb{\mu}|\pmb{\mu}\rangle}\bigg|_{\text{1ph}}\bigg)^{-1}.
\ee
For the same states we have also checked the accuracy of the exponentiations \fr{resum1}, \fr{resum2}.
Results for $L=N=128$ and $c=100$ are shown in Table \ref{tab:FFcomp}
\begin{table}[ht]
\begin{tabular}{|c|c|c|c|c|c|c|c|}
\hline
P&$2J_a$&$2I_a$&${A^{(r)}_1}/{A_1}$&$A^{\rm (r)}_2/A_2$& $A^{\rm (r)}_3/A_3$ & $A^{\rm (r)}_6/A_6$& $R$\\
\hline
-5.15  &-287   & -77     &0.998   &1.00    &1.03    &0.995   &1.022\\ \hline
-3.14  &-261   & -133    &0.999   &1.00    &1.02    &0.991   &1.014\\ \hline
5.35   &223    & 5       &0.999   &1.00    &1.02    &1.      &1.018\\ \hline
-4.47  &-173   & 9       &1.00    &1.00    &1.01    &1.      &1.013\\ \hline
4.57   &207    & 21      &0.999   &1.00    &1.01    &1.00    &1.015\\ \hline
-2.85  &-155   & -39     &1.00    &1.00    &1.01    &1.      &1.009\\ \hline
-6.92  &-243   & 39      &0.999   &1.00    &1.02    &1.      &1.025\\ \hline
6.97    &273   & -11     &0.998   &1.00    &1.02    &1.00    &1.027\\ \hline
-6.38  &-225   & 35      &0.999   &1.00    &1.02    &1.      &1.021\\ \hline
-4.61  &219    &-31     &0.999   &1.00    &1.02    &1.      &1.016\\ \hline
\end{tabular}
\caption{Results for the ratios of the resummed $1/c$-expansions
for matrix elements to the numerically exact results for $c=100$,
 $L=N=128$, $\beta=1$. We see that the resummed expression works
 rather well. As expected it does become worse when the excited state
 involves larger differences $|\lambda_p-\lambda_h|$ as the
 $1/c$-expansion requires these to be small compared to $c$.} 
\label{tab:FFcomp}
\end{table}
We see that the results are quite satisfactory. The values of $\beta$
and $c$ have been chosen to ensure that the differences
$|\lambda_{j,k}|$ are all small compared to $c$, which is a key
assumption of the $1/c$-expansion, \emph{cf.} the discussion in
Ref.~\cite{granet2020a}. We note that the factors 
$A^\text{(r)}_n$ are generally quite different from their "bare"
values, which indicates the breakdown of the bare
$1/c$-expansion. However, in the final expression \fr{ff1phres} a
number of cancellations occur, which render the resummed result
\fr{ff1phres} very close to the bare expression \fr{ff1ph3} for the
parameters considered here.

\section{Sampling macro-states in a finite volume}
\label{app:sampling}
In this appendix we discuss in some detail how to sample a given
macro-state for a large, finite number of particles. The key element
is to generate appropriate sets of ``Bethe integers'' $I_j$, which
characterize the solutions of the Bethe equations \fr{BAE2}. For
simplicity we focus on the impenetrable case $c=\infty$, where 
\begin{align}
\lambda_j=\frac{2\pi I_j}{L}\ .
\end{align}
Thermal states are of particular interest as they are the most
abundant states at a given energy density and we therefore focus on
them in our discussion. The generalization to atypical finite entropy
states is straightforward. To be specific we take $c=\infty$ and
consider a temperature $T=10$ and chemical potential
$\mu=12.1058$. This gives particle density $D=1$, energy density
\be
e_\infty=8.036608362699118\ ,
\ee
and a root density
\be
\rho(x)=\frac{1}{2\pi}\frac{1}{1+e^{(x^2-\mu)/T}}\ .
\ee
The corresponding density of $I_j/L$ is simply
\be
\varrho(\nu)=\frac{1}{1+e^{((2\pi \nu)^2-\mu)/T}}\ .
\label{varrho}
\ee
For later convenience we define a cumulative probability distribution
function 
\be
C(\nu)=\frac{1}{D}\int_0^\nu d\nu'\ \varrho(\nu')\ .
\label{cumu}
\ee
\subsection{Micro-canonical ensemble}
\begin{figure}[ht!]
\begin{center}
\includegraphics[scale=0.5]{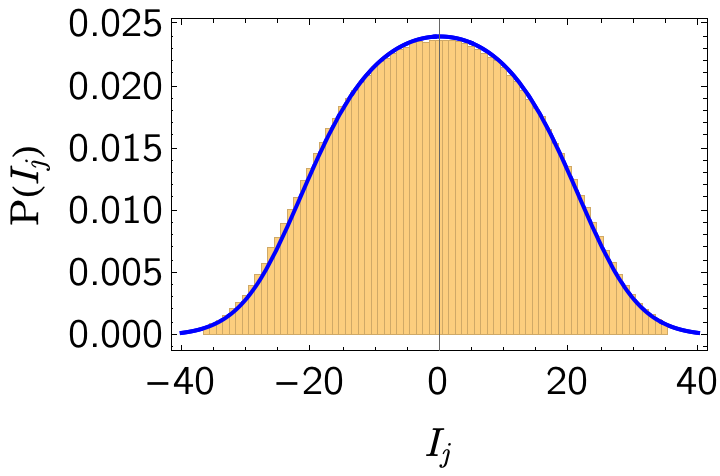}
\end{center}
\caption{Histogram of integers occurring in the micro-canonical
ensemble \fr{EMC}. The blue curve is the thermodynamic
root distribution function $\frac{1}{L}\varrho(\frac{I}{L})$.
}
\label{fig:L=32}
\end{figure}

\begin{figure}[ht!]
\begin{center}
\includegraphics[scale=0.5]{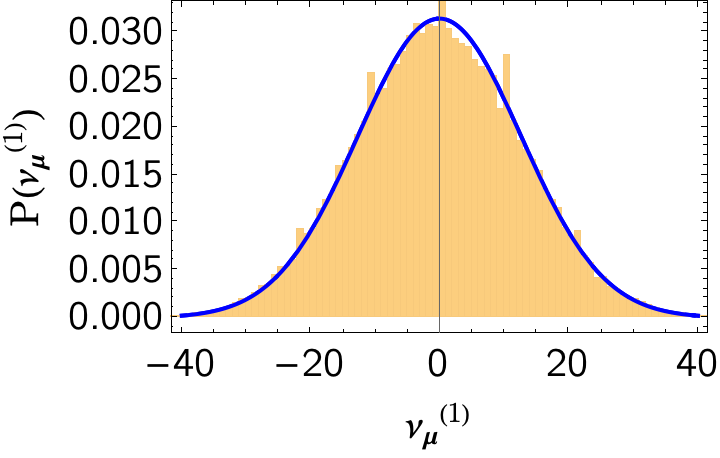}
\end{center}
\caption{Histogram of the total momentum 
in the micro-canonical ensemble \fr{EMC}. The curve is a fit to a
normal distribution.}
\label{fig:mtm_MC}
\end{figure}

\begin{figure}[ht!]
\begin{center}
\includegraphics[scale=0.5]{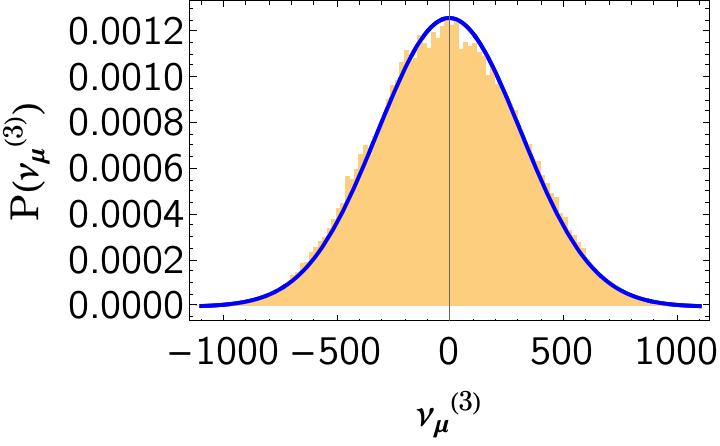}
\end{center}
\caption{Histogram of the eigenvalues
of the third conservation law in the micro-canonical ensemble
\fr{EMC}. The curve is a fit to a normal distribution.}
\label{fig:q3_MC}
\end{figure}

Let's start by randomly sampling distinct integers and just fixing an
energy window for ``acceptable states''. We now fix our particle
number and system size to be $N=L=32$ and consider energies in the
window
\be
|E-e_\infty L|<2\ .
\label{EMC}
\ee
In order to be able to sample energy eigenstates we also need to
impose the constraint 
\be
|I_j|<I_{\rm max}\ ,
\label{integercutoff}
\ee
where the values of $I_{\rm max}$ we have considered are $I_{\rm
  max}\leq 36$. The cutoff \fr{integercutoff} is required as the
numerical cost for finding configurations that fulfil \fr{EMC}
increases exponentially with $N$.
The histogram of integers occurring in eigenstates fulfilling
this constraint is shown in Fig.~\ref{fig:L=32}.
We see that our micro-canonical ensemble nicely reproduces the
thermodynamic root density \fr{varrho}. In Figs~\ref{fig:mtm_MC} and
\ref{fig:q3_MC} we show results for the probability distributions of
total momentum and the  third conservation law  
\be
\nu^{(1)}_{\bla}=\sum_{n=1}^{32}\lambda_n\ ,\qquad
\nu^{(3)}_{\bla}=\sum_{n=1}^{32}\lambda_n^3\ 
\ee
in the micro-canonical ensemble.

\begin{figure}[ht!]
\begin{center}
\includegraphics[scale=0.5]{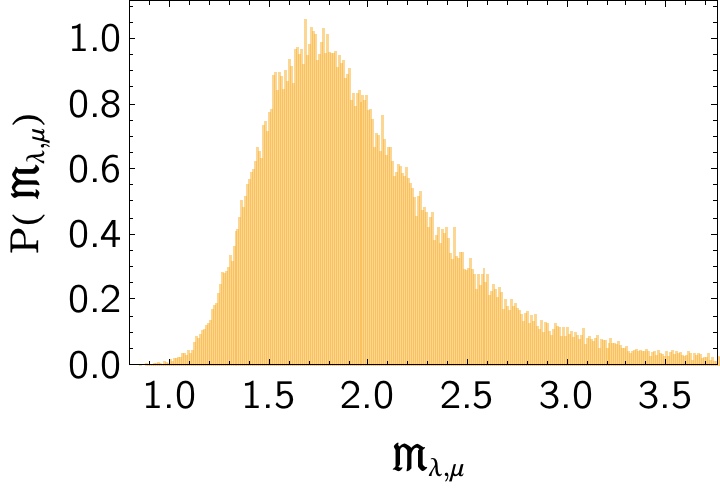}
\end{center}
\caption{Histogram of the logarithm of matrix elements
  $\mathfrak{M}_{\bla,\bmu}$ \fr{MEfrak} between the smooth ket state
  and bra states obtained by micro-canonical sampling \fr{EMC}. }
\label{fig:field_MC}
\end{figure}

The averages for these conserved quantities are as expected zero, but
the spread of eigenvalues is very large. Finally we show the
probability distribution of the matrix elements of the Bose field
$\mathfrak{M}_{\bla,\bmu}$ \fr{MEfrak} in Fig.~\ref{fig:field_MC}
The micro-canonical sampling described here cannot be used for
large particle numbers because it is extremely inefficient. The set of
half-odd integer numbers we need to sample has dimension
\be
\binom{2I_{\rm max}}{N}\ ,
\ee
which grows exponentially with the number of particles. We therefore
require more efficient ways of sampling the relevant micro-states.

\subsection{Plain vanilla box sampling (PVBS)}
\begin{figure}[ht!]
\begin{center}
\includegraphics[scale=0.5]{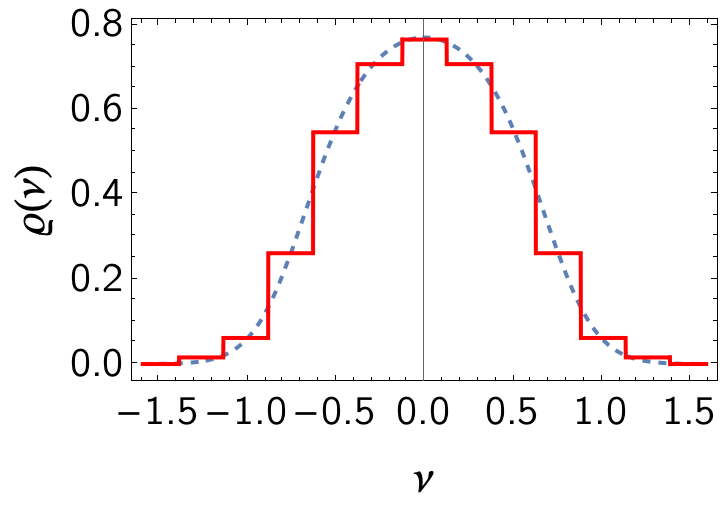}\qquad
\end{center}
\caption{Discretization of the distribution function $\varrho(\nu)$.}
\label{fig:rhoD}
\end{figure}

\begin{figure}[ht!]
\begin{center}
\includegraphics[scale=0.5]{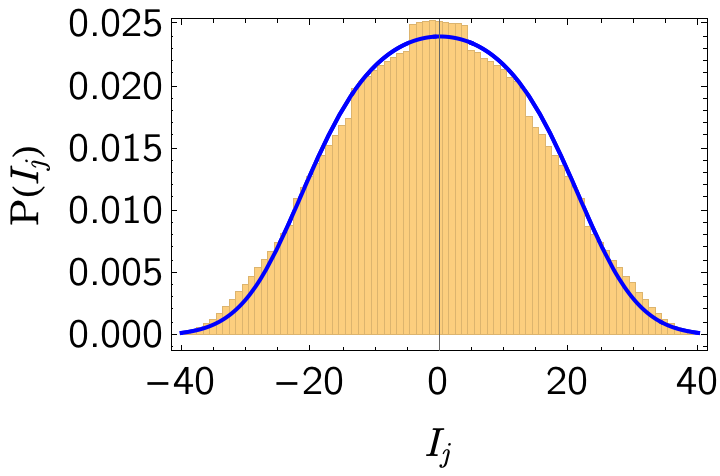}
\end{center}
\caption{Histogram of integers obtained by PVBS with $M=7$. The blue
  curve is the thermodynamic root distribution function
  $\frac{1}{L}\varrho(\frac{I}{L})$. 
}
\label{fig:L=32_7bin}
\end{figure}

The simplest idea for targeting energy eigenstates in the appropriate
energy window is to use ``box-sampling'' of the probability distribution
$\varrho(\nu)/D$. The rationale behind this is that for very large
numbers of particles almost all these states will correspond to a
discretization of $\varrho(\nu)$, \emph{cf.} the steps leading to our 
expression for the entropy \fr{entropy}.
So what one would do is to approximate $\varrho(\nu)$ as shown in
Fig.~\ref{fig:rhoD}. 
To that end we introduce a cut-off $a_0$ (which corresponds to $-I_{\rm
  max}$ in the discussion of the micro-canonical ensemble in the
previous subsection). 
and then sub-divide the $\nu$ axis into
intervals $B_n=[a_{n-1},a_n]$ for $1\leq n\leq M$. The density of
$\nu_j=I_j/L$ in $B_n$ is then taken to be 
\be
\varrho_n\equiv D[C(a_n)-C(a_{n-1})],
\ee
where the cumulative PDF $C(\nu)$ is defined in \fr{cumu}. We can
straightforwardly translate this into a distribution function of
(half-odd) integers that is piecewise constant on the $M$ intervals
\be
\bar{B}_n=[J_{n-1},J_n).
\ee
The number of integers in box $\bar{B}_n$ is $J_n-J_{n-1}$, where
in practice we adjust $a_0$ in such a way that
\be
\sum_{n=1}^MN_n=N=DL.
\ee
We now sample the discretized probability distribution box-by-box:
from the (half-odd) integers in $\bar{B}_n$ we randomly select $N_n$
elements, where
\be
N_n=\text{Round}\Big(L\varrho_n\Big).
\ee
The total number of different configurations in the resulting sample
space is
\be
\mathfrak{N}_M(\{N_j\})=\prod_{n=1}^M{{I_n-I_{n-1}}\choose{N_n}}\equiv
e^{L\mathfrak{s}_M}\ .
\ee
This is much smaller than the number of configurations that needs to
be sampled in the micro-canonical ensemble discussed earlier, which
corresponds to the choice $M=1$. Importantly, in the double limit
\be
\lim_{M\to\infty}\lim_{L\to\infty}\mathfrak{s}_M
\ee
the number of micro-states produced by this procedure recovers the
correct entropy density of the thermal macro-state under
consideration. One may therefore expect that this procedure provides a
good way of sampling thermal states in finite systems. For a finite
number of particles the number of sampled states decreases with $M$
and in our example we find 
\be
\mathfrak{s}_3=1.04\ ,\quad
\mathfrak{s}_5=1.01\ ,\quad
\mathfrak{s}_7=0.952773.
\ee
These values should be compared with the thermodynamic result
$1.20041$. In practice we still need to impose the energy-window
restriction \fr{EMC} so that the actual numbers of states are
smaller. For the finite particle numbers of relevance here PVBS does
not agree well with the micro-canonical sampling. To show the degree
of difference we present results for $M=7$ and our $L=N=32$ example.
In Fig.~\ref{fig:L=32_7bin} we show the distribution of integers,
which reproduces the thermodynamic root distribution function
in a satisfactory manner.

\begin{figure}[ht!]
\begin{center}
\includegraphics[scale=0.5]{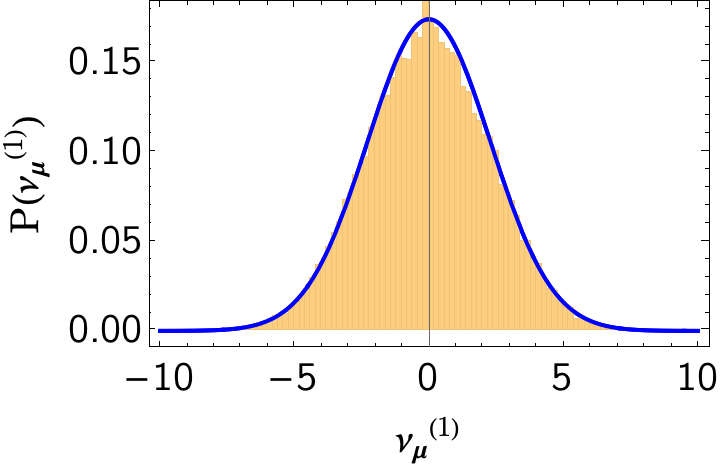}
\end{center}
\caption{Histogram of the total momentum obtained by PVBS with $M=7$
The curve is a fit to a normal distribution.}
\label{fig:mtm_7bin}
\end{figure}

\begin{figure}[ht!]
\begin{center}
\includegraphics[scale=0.5]{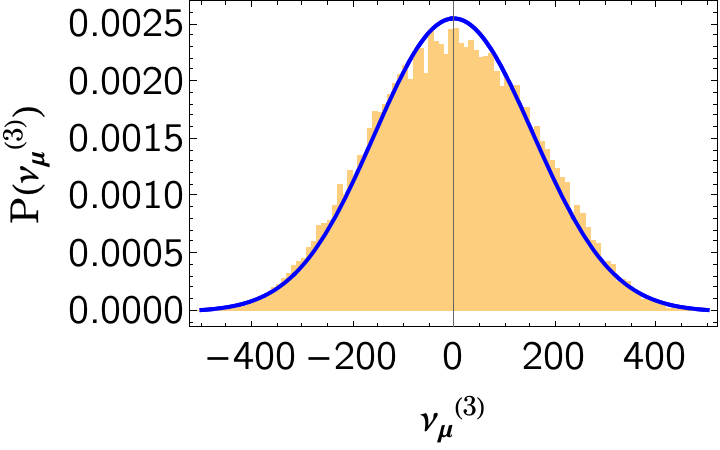}
\end{center}
\caption{Histogram of the eigenvalues of the third conservation law obtained by PVBS with $M=7$
The curve is a fit to a normal distribution.}
\label{fig:q3_7bin}
\end{figure}

\begin{figure}[ht!]
\begin{center}
\includegraphics[scale=0.5]{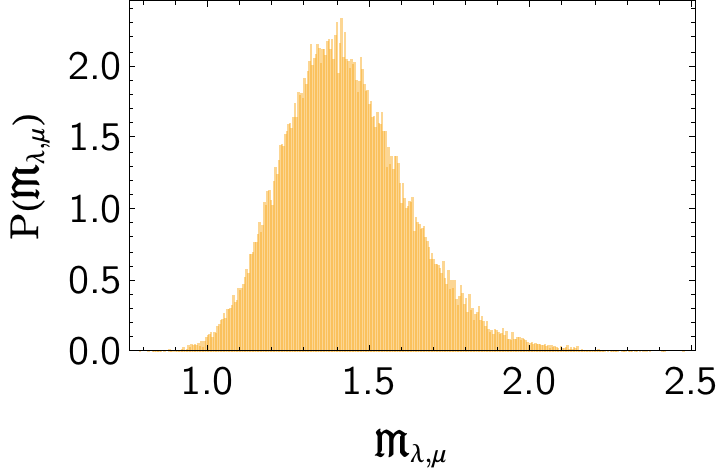}
\end{center}
\caption{Histogram of the logarithm of matrix elements
  $\mathfrak{M}_{\bla,\bmu}$ \fr{MEfrak} between the smooth ket state
  and bra states obtained by PVBS with $M=7$. }
\label{fig:field_7bin}
\end{figure}

In Figs~\ref{fig:mtm_7bin} and \ref{fig:q3_7bin} we show the
distribution of the total momentum and third conservation laws
respectively. 
We observe that both distributions are very considerably narrower than
the corresponding ones for micro-canonical sampling, see
Figs~\ref{fig:mtm_MC} and \ref{fig:q3_MC}. Finally we show the
probability distribution of the matrix elements of the Bose field
$\mathfrak{M}_{\bla,\bmu}$ \fr{MEfrak} in Fig.~\ref{fig:field_7bin}.
We observe that the typical matrix elements obtained by PVBS are
significantly larger than in micro-canonical sampling, \emph{cf.}
Fig.~\ref{fig:field_MC}.
The differences in the probability distributions of matrix elements
and the eigenvalues of conserved quantities between PVBS and MC
sampling is easy to understand intuitively: by construction PVBS
produces significantly smaller fluctuations that the MCE in finite
volumes. While the expectation is that these finite-size effects will
disappear as the thermodynamic limit is approached, they severely
limit the utility of PVBS for the (numerically) accessible system sizes.
\subsection{Fluctuating box sampling}
\begin{figure}[ht!]
\begin{center}
\includegraphics[scale=0.45]{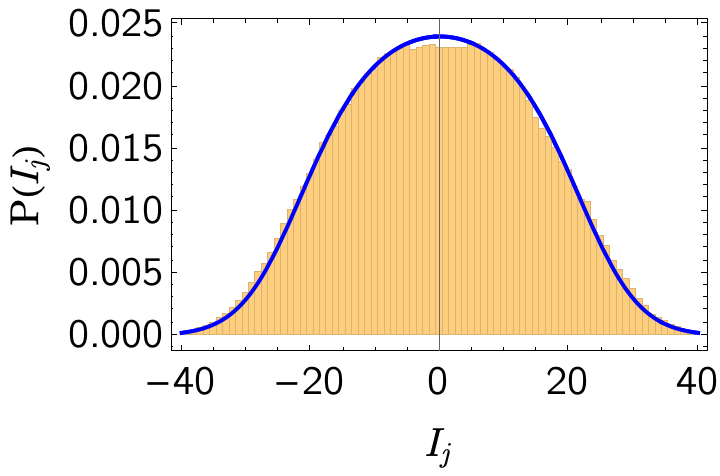}\quad
\end{center}
\caption{Histograms of integers produced by FBS with $M=7$ and energies
in the window $|E-e_\infty L|<2$.}
\label{fig:L=32F7box}
\end{figure}

\begin{figure}[ht!]
\begin{center}
\includegraphics[scale=0.45]{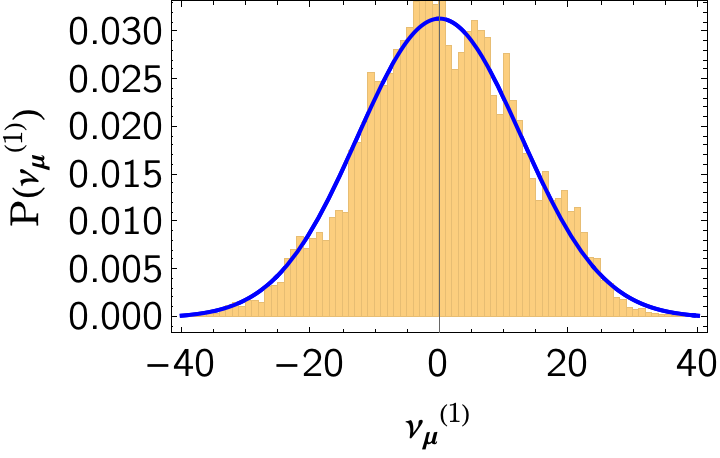}
\end{center}
\caption{Histograms of momentum produced
by FBS with $M=7$ compared to the MCE (solid blue line).}
\label{fig:L=32F7boxb}
\end{figure}

\begin{figure}[ht!]
\begin{center}
\includegraphics[scale=0.45]{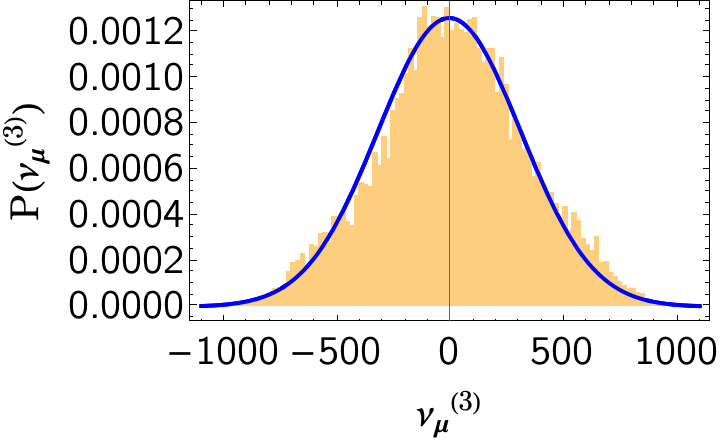}
\end{center}
\caption{Histograms of $\nu^{(3)}_{\bla}$ produced
by FBS with $M=7$ compared to the MCE (solid blue line).}
\label{fig:L=32F7boxc}
\end{figure}

\begin{figure}[ht!]
\begin{center}
\includegraphics[scale=0.45]{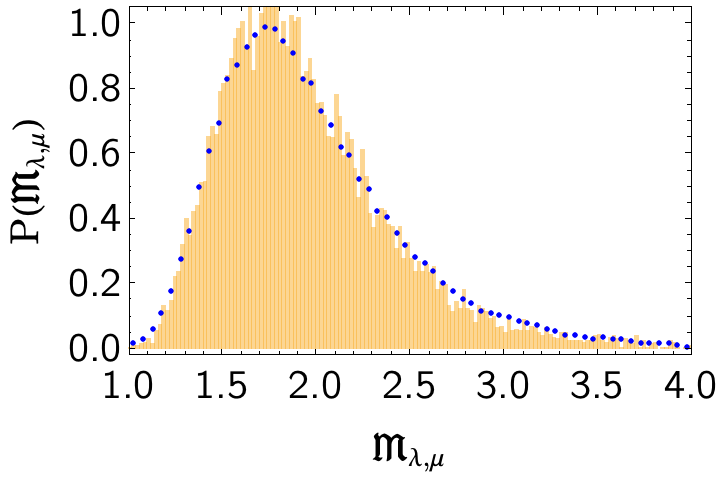}
\end{center}
\caption{Histogram of the matrix elements between the smooth ``ket''
state and energy eigenstates in the window $|E-e_\infty L|<2$ obtained
by FBS (yellow) and in the micro-canonical ensemble (blue dots).}
\label{fig:L=32MEFBS}
\end{figure}

As we have seen, in mesoscopic volumes the PVBS accesses a much more
restrictive set of energy eigenstates than the MCE. We can make up for
this by allowing the box occupation numbers $N_n$ to fluctuate. Given
a set $B$ of boxes with vacancies ($V_1,\dots,V_M$) we generate a set
of occupation numbers $\{N_j\}$ such that
\be
\sum_jN_j=N\ ,
\ee
where we allow the $N_j$ to fluctuate as follows. Let
$N^{(\text{PV})}_j$ be the PVBS particle numbers. We then take 
\be
N_j=N^{(\text{PV})}_j+\delta N_j\ ,
\ee
where the random integers $\delta N_j$ are taken to add up to zero and fulfil
\be
\delta N_j={\cal O}\big(\text{min}\Big\{\sqrt{N^{(\text{PV})}_j},\sqrt{|V_j-N^{(\text{PV})}_j|}\Big\}\big).
\ee
Given a set of particle numbers ($N_1,\dots,N_M$) we calculate the number of
micro-states obtained by box-sampling 
\be
\mathfrak{N}_M(\{N_j\},\{V_j\})=\prod_{n=1}^M{{V_n}\choose{N_n}}\ .
\ee
We then generate
\be
\bigg[\frac{\mathfrak{N}_M(\{N_j\},\{V_j\})}{N_0}\bigg]
\ee
samples from the configuration specified by $\{N_1,\dots,N_M\}$, where
$N_0$ is some fixed reference number. By construction this procedure
increases fluctuations. In practice we may choose the outermost boxes
to be larger in order to decrease ``tail effects''. Results obtained
by this method for $N=L=32$ are shown in
Figs~\ref{fig:L=32F7box}, \ref{fig:L=32F7boxb}, \ref{fig:L=32F7boxc} and \ref{fig:L=32MEFBS}.

We see that the fluctuating box sampling reproduces the results of the
MCE fairly well. However, this requires the fluctuations are taken to be
sufficiently strong. In particular, if we make them weaker by changing
the coefficient that multiplies the r.h.s. in the expression for
$\delta N_j$ the agreement becomes worse. This is as expected. FBS is
significantly slower than PVBS and becomes computationally very
expensive for large particle numbers.

\subsection{Random sampling (RS)}
\begin{figure}[ht!]
\begin{center}
\includegraphics[scale=0.45]{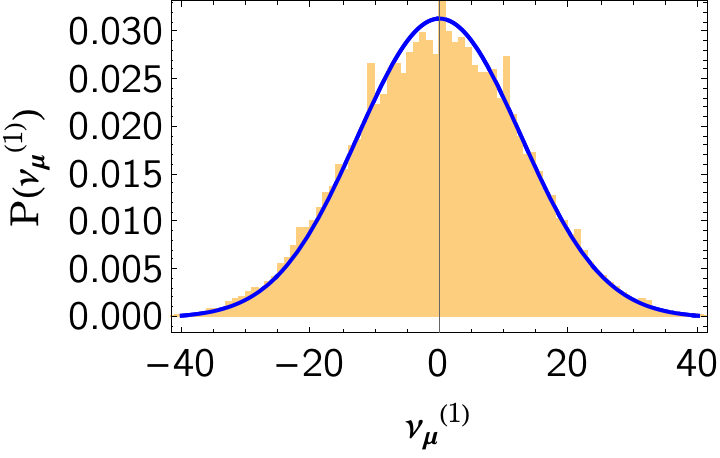}
\end{center}
\caption{Histograms of the total momentum obtained by random sampling
under the constraint that $|E-e_\infty L|<2$ for $N=L=32$. The results
in the micro-canonical ensemble are shown as the solid blue line.} 
\label{fig:L=32RSmtm}
\end{figure}

\begin{figure}[ht!]
\begin{center}
\includegraphics[scale=0.45]{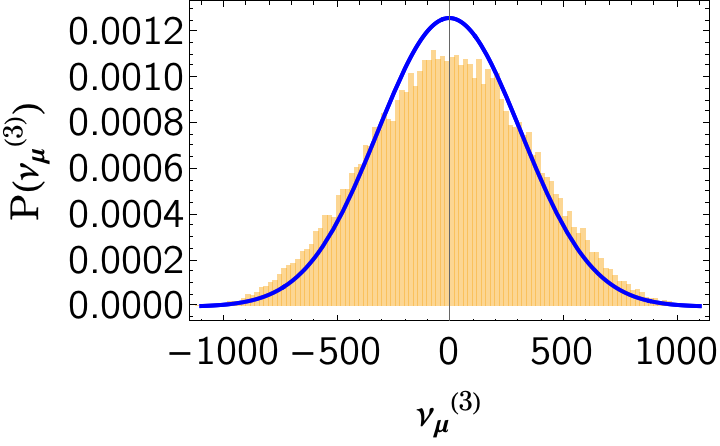}
\end{center}
\caption{Histograms of the third conservation law obtained by random sampling
under the constraint that $|E-e_\infty L|<2$ for $N=L=32$. The results
in the micro-canonical ensemble are shown as the solid blue line.} 
\label{fig:L=32RSq3}
\end{figure}

\begin{figure}[ht!]
\begin{center}
\includegraphics[scale=0.45]{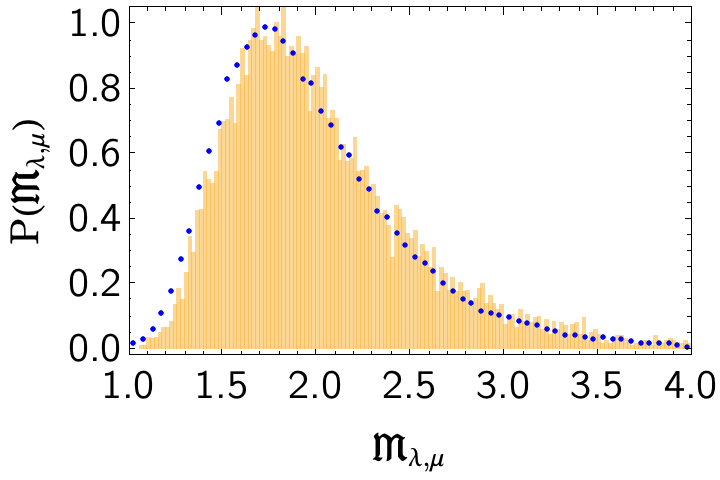}
\end{center}
\caption{Histograms of the matrix elements of the Bose field operator
  between the smooth ``ket'' state and energy eigenstates in the
  window $|E-e_\infty L|<2$ for $N=L=32$ obtained by random sampling (yellow) and
  in the micro-canonical ensemble (blue dots).} 
\label{fig:L=32RSME}
\end{figure}

The distribution of integers in the micro-canonical ensemble is
well-described by the thermodynamic distribution function
$\varrho(\nu)/D$. This suggests that a random sampling of this probability
distribution should reproduce the MCE. The difficulty is that we must
generate \emph{non-repeating} integers. Our starting point is the set
of integers
\be
S_M=\{j|-I_{\rm max}\leq j\leq I_{\rm max}\}\ , \quad M=2I_{\rm
max}+1.
\label{SM}
\ee
and an associated discrete probability distribution
\be
P_M=\{p_1,\dots, p_M\}\ .
\ee
In practice we take $P_M$ to be a discretization of self-consistently
determined continuous PDF $P(\nu)$. The corresponding set of
cumulative probabilities is 
\be
{\cal C}_M=\{C_n=\sum_{j=1}^{n-1}p_j\ ,\quad n=1,\dots, 2I_{\rm max}+2\}.
\ee
We now generate a (real) random number $r$ in the interval $[0,1]$ and
determine the integer $j$ such that
\be
C_{j-1}<r<C_j\ .
\ee
We then remove the integer $j$ from the set $S_M$ and define a new
discrete probability distribution
\be
P_{M-1}=\Big\{\frac{p_1}{1-p_j},\dots,\frac{p_{j-1}}{1-p_j},\frac{p_{j+1}}{1-p_j},\dots,
\frac{p_M}{1-p_j}\Big\}\ 
\ee
and the associated cumulative probability distribution ${\cal
  C}_{M-1}$.
Repeating this procedure $N$ times results in a set of
distinct integers $\{I_1,\dots,I_N\}$. Finally, we impose that the
probability distribution of these sets of integers is a discretization
of the (normalized) root density $\varrho(\nu)/D$. Importantly, this
requires an initial probability distribution $P(\nu)$ that is different
from $\varrho(\nu/D)$.
The PDF required to produce the normalized root distribution upon random
sampling is shown in the main text in Fig.~\ref{fig:priort10}.
\begin{figure}[ht!]
\begin{center}
\includegraphics[scale=0.45]{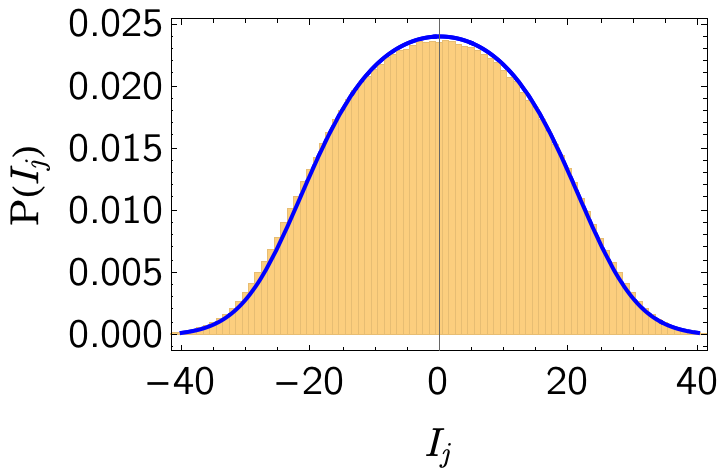}
\end{center}
\caption{Histogram of integers by RS using the probability
distribution $P(\nu)$ shown in Fig.~\ref{fig:priort10}. The solid
curve is the root distribution function $\varrho(\nu)/D$.} 
\label{fig:rhoren}
\end{figure}
In Figs~\ref{fig:L=32RSmtm}, \ref{fig:L=32RSq3} and \ref{fig:L=32RSME}
we show the histograms obtained by our random sampling procedure for
the eigenvalues of momentum, the third conservation law and
the matrix elements of the Bose field operator between the smooth
``ket'' state and energy eigenstates in the window $|E-e_\infty L|<2$.

We observe that the results are in good agreement with those obtained
by micro-canonical sampling. We conjecture that the remaining
differences, in particular in $P(\nu^{(3)}_{\bmu}$, are at least
partially caused by the cutoff in the MC sampling procedure.  
\subsection{Simplified random sampling (SRS)}
\begin{figure}[ht!]
\begin{center}
\includegraphics[scale=0.45]{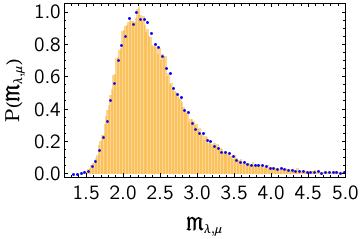}
\end{center}
\caption{Histogram the matrix elements of the Bose field operator
between the smooth ``ket'' state and energy eigenstates in the
window $|E-e_\infty L|<10$ obtained by SRS (yellow) and analogous
result for RS (blue dots).} 
\label{fig:L=128RS}
\end{figure}

\begin{figure}[ht!]
\begin{center}
\includegraphics[scale=0.45]{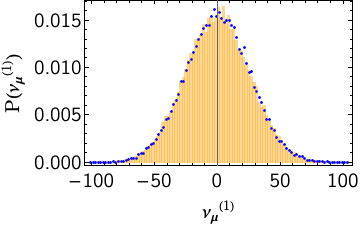}
\end{center}
\caption{Histogram of the eigenvalues of momentum for
energy eigenstates in the window $|E-e_\infty L|<10$ obtained
by SRS (yellow) and analogous result for RS (blue dots).}
\label{fig:L=128RS_mtm}
\end{figure}

\begin{figure}[ht!]
\begin{center}
\includegraphics[scale=0.45]{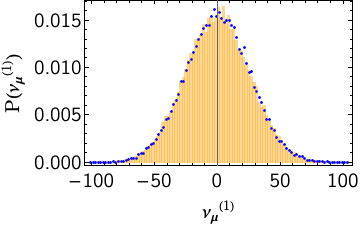}
\end{center}
\caption{Histogram of the eigenvalues of the third conservation law
for energy eigenstates in the window $|E-e_\infty L|<10$ obtained
by SRS (yellow) and analogous result for RS (blue dots).}
\label{fig:L=128RS_q3}
\end{figure}

The random sampling algorithm described above is somewhat slow. We
therefore use the simplified algorithm described in 
section~\ref{sec:sampling} of the main text. The latter is faster as
it treats the constraint that all (half-odd) integers must be distinct
in a much simpler fashion. It nevertheless gives results that agree with
RS within the statistical error in all cases we have tested. Examples
are shown in Figs~\ref{fig:L=128RS}, \ref{fig:L=128RS_mtm} and
\ref{fig:L=128RS_q3}. Here we have chosen a larger energy window
$|E-e_\infty L|<10$.

\subsection{Interacting case}
As all sampling methods discussed above are based on drawing sets
of non-repeating (half-odd) integers $\{I_j\}$ from a probability distribution
they generalize in a straightforward way to the interacting case. The
main differences are:
\begin{itemize}
\item{} The target PDF $P(I_j)$ is obtained from \fr{varrho_thermal} by
solving the nonlinear integral equations \fr{TBAeqs} (for thermal
macro-states). 
\item{} For $0<c<\infty$ we need to (numerically) solve the Bethe
  equations once we have generated a set $\{I_j\}$.
\end{itemize}

\bibliography{bibliography}

\end{document}